\documentclass{article}

\usepackage{arxiv}

\usepackage[utf8]{inputenc} % allow utf-8 input
\usepackage[T1]{fontenc}    % use 8-bit T1 fonts
\usepackage{hyperref}       % hyperlinks
\usepackage{url}            % simple URL typesetting
\usepackage{booktabs}       % professional-quality tables
\usepackage{amsfonts}       % blackboard math symbols
\usepackage{nicefrac}       % compact symbols for 1/2, etc.
\usepackage{microtype}      % microtypography
\usepackage{lipsum}		% Can be removed after putting your text content
\usepackage{graphicx}
\usepackage{doi}
\usepackage{enumitem}
\hypersetup{
pdftitle={Cutting soft materials: how material differences shape the response},
pdfsubject={cond-mat.mtrl-sci},
pdfauthor={Miguel Angel Moreno-Mateos, Paul Steinmann},
pdfkeywords={Soft materials, Soft fracture mechanics, Contact mechanics, Computational mechanics, Experimental mechanics, Food mechanics},
}

\hypersetup{
    colorlinks=true,
    linkcolor=blue,     % Color for internal links
    filecolor=magenta,  % Color for file links
    urlcolor=black,      % Color for URLs
    citecolor=blue,    % Color for citations
}

\usepackage{booktabs}
\usepackage{amssymb}
\usepackage{amsmath}
\usepackage{stmaryrd}
\usepackage{mathrsfs}
\usepackage{color}
\usepackage{graphicx}
\usepackage{latexsym}
\usepackage{tabls}
\usepackage{graphicx}
\usepackage{epsfig}
\usepackage{subfigure}
\usepackage{rotating}
\usepackage{latexsym}
\usepackage[mathcal]{eucal}
\usepackage{amssymb}
\usepackage{colortbl}
\usepackage{times}
\usepackage{longtable}
\usepackage{algorithmic}
\usepackage{algorithm}
\usepackage{psfrag}
\usepackage{tabularx}
\usepackage{epstopdf}
\usepackage{gensymb}
\usepackage{bbm}
\usepackage{setspace}
\usepackage{fullpage}
\usepackage{url}
\usepackage{calc}
\usepackage{siunitx}

\usepackage{lineno,hyperref}
\usepackage{stmaryrd}
\usepackage{subfigure}
\modulolinenumbers[50]
\usepackage{epstopdf}
\usepackage{array}
\usepackage{textcomp}

\usepackage{booktabs}
\usepackage{mathrsfs}
\usepackage{multirow}
\usepackage{lscape}
\usepackage{amsmath}

\usepackage{mathtools}
\usepackage{amssymb}
\usepackage{color,soul}
\usepackage{bm}

\usepackage{titlesec}
\titleformat{\paragraph}
{\normalfont\normalsize\bfseries}{\theparagraph}{1em}{}
\titlespacing*{\paragraph}
{0pt}{3.25ex plus 1ex minus .2ex}{1.5ex plus .2ex}

\usepackage{tikz}

\usepackage{makecell}
\usepackage{bbm}
\usepackage{float}

\usepackage{amsthm}
\theoremstyle{definition}
\newtheorem*{remark}{Remark}

\usepackage{lineno}

\usepackage[font=scriptsize]{caption}
\begin{document}

\title{Cutting soft materials: how material differences shape the response}

\author{ \href{https://orcid.org/0000-0002-3476-2180}{\includegraphics[scale=0.06]{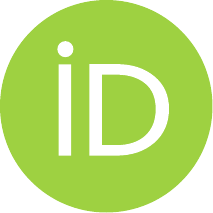}\hspace{1mm}Miguel Angel Moreno-Mateos}\thanks{Corresponding author.} \\
	Institute of Applied Mechanics\\
	Friedrich-Alexander-Universität Erlangen–Nürnberg\\
	91058, Erlangen, Germany\\
	\texttt{miguel.moreno@fau.de} \\
	%% examples of more authors
	\And
	\href{https://orcid.org/0000-0003-1490-947X}{\includegraphics[scale=0.06]{orcid.pdf}\hspace{1mm}Paul Steinmann} \\
	Institute of Applied Mechanics\\
	Friedrich-Alexander-Universität Erlangen–Nürnberg\\
	91058, Erlangen, Germany\\
	Glasgow Computational Engineering Centre\\
	University of Glasgow\\
	G12 8QQ, United Kingdom\\
}

\maketitle

\begin{abstract}
Cutting soft materials is a complex process governed by the interplay of bulk large deformation, interfacial soft fracture, and contact forces with the cutting tool. Existing experimental characterizations and numerical models often fail to capture the variety of observed cutting behaviors, especially the transition from indentation to cutting and the roles of dissipative mechanisms. Here, we combine novel experimental cutting tests on three representative materials---a soft hydrogel, elastomer, and food-based materials---with a coupled computational model that integrates soft fracture, adhesion, and frictional interactions. Our experiments reveal material-dependent cutting behaviors, with abrupt or smooth transitions from indentation to crack initiation, followed by distinct steady cutting regimes. The computational model captures these behaviors and shows that adhesion and viscous cohesive forces dominate tangential stresses, while Coulomb friction plays a negligible role due to low contact pressures. Together, these results provide new mechanistic insights into the physics of soft cutting and offer a unified framework to guide the design of soft materials, cutting tools, and cutting protocols, with direct relevance to surgical applications and the engineering of food textures optimized for mastication.
\end{abstract}


\section{Introduction}
Cutting is a fundamental process in both nature and technology, from insects slicing through plants \cite{Puffel2025c}, package opening \cite{Pagani2015}, metal, rock, and wood cutting \cite{Jones1987,Nishimatsu1972,Jocic2025}, to robotic tools dissecting soft tissues \cite{Takabi2017} or food products \cite{Xu2022cutting}. Yet, when it comes to soft solids, the mechanics of cutting remain poorly understood. Unlike brittle or hard materials, soft solids exhibit large deformations, rate-dependent dissipation, frictional sliding, adhesion, and even wear at the tool interface, all of which shape the transition from indentation to fracture \cite{Terzano2018,Goda2024b}. Despite the prevalence of soft cutting in areas such as biomedical engineering \cite{Chanthasopeephan2003,Hu2013cutting,Brodie2018,Bui2019}, food processing \cite{Kamyab1998,Wright2024}, and soft robotics, a general framework capturing the physics of soft cutting is still missing.

Understanding this process raises several open questions: How does the initial indentation relate to the onset of cutting? What is the role of the material’s constitutive behavior in determining the cutting force? How do adhesion, wear, and Coulomb friction compete in controlling the cutting resistance? These questions challenge traditional models of fracture and friction. Classical fracture mechanics \cite{Williams2010} and cohesive zone models typically assume sharp crack tips and separation governed by stress intensity or traction-separation laws. Yet, in cutting soft solids, experiments show that failure may initiate under a blunted blade through a progressive, geometry- and material-dependent process \cite{Gent1996,McCarthy2007a,McCarthy2007b,Zhang2019cutting,Zhang2021cutting}. Together with numerical simulations, recent studies demonstrated that cutting produces a stronger strain localization than fracture \cite{Mars2019,Robertson2021}. This process has been also connected to the microstructure in elastomer \cite{Zhao2025c} and rubber materials \cite{Nian2025}. Recent studies describe the fracture mechanics of polymer and composite structures across scales and related design pathways to enhance their failure response \cite{Creton2016,Slootman2022,Liu2023,Chen2020p,Tian2025}. The volumetric response of polymer networks has been outlined and may be key in soft cutting mechanics \cite{Hartquist2024}. Analytic results connect stress intensity factors with tip sharpness and predict whether a crack propagates autonomously or remains in contact with the cutting tool. Computational and theoretical studies have begun to explore cohesive fracture and crack evolution in soft solids under localized loading \cite{Spagnoli2018,Russ2020,Moreno-Mateos2024b}, laying important groundwork for understanding cutting as a soft fracture process.

The transition from stable indentation to unstable soft fracture is not purely governed by bulk toughness but modulated by viscous dissipation, adhesive interactions at the tool–material interface \cite{Zhou2024b}, and material compressibility \cite{Goda2025}. Although experimental setups have been proposed to investigate cutting in the absence of friction \cite{Lake1980,Zhan2024cutting}, this is not representative of most cutting processes where a cutting tool remains in contact with the material as it penetrates it \cite{Triki2017,Liu2021cutting2}. Moreover, materials of similar bulk stiffness can behave very differently under cutting. Moist foodstuffs, for instance, exhibit cutting forces inconsistent with purely elastic or brittle theories; instead, rate-dependent and viscous failure modes are more accurate descriptors \cite{Cho1998,Danas2009,Schuldt2016,Boisly2016}. Double-network hydrogels exhibit extreme cutting toughness \cite{Gong2003cutting}. Cryogenic cutting of elastomers alters their constitutive behavior to enhance cuttability, effectively shifting the material response toward conditions optimal for cutting \cite{Maurya2021}. This raises fundamental questions about how energy is partitioned between elastic storage, fracture creation, and dissipative pathways—questions that traditional theories cannot fully answer.

In this work, we combine cutting experiments and a numerical model based on a cohesive zone with contact approach to investigate how soft materials fail under a moving blade. Our model mimics three-dimensional cutting experiments explicitly accounting for adhesion, friction, and viscous dissipation. This provides a mechanistic framework that goes beyond classical approaches. In particular, our unified framework substantiates specific hypotheses that lead to four central findings: the material and dissipative parameters directly modulate unstable/stable cutting transition; cutting initiates in the center beneath the blade; adhesion, rather than Coulomb friction, governs tangential forces during cutting; and cutting of moist foodstuffs can be explained by viscous dissipation rather than fracture energy. 

Together, these findings offer a new understanding of soft cutting as a coupled problem of material response, interface mechanics, and energy dissipation. Our results clarify why nominally similar materials respond differently to the same blade and provide general principles applicable to biological, synthetic, and food-based soft solids. By bridging experiment and modeling, we propose a unified approach to a classically fragmented problem—shedding light on one of the most common yet least understood physical processes in soft matter.

\newpage

\section{Results and Discussion}

\subsection{Material behavior, fracture, and friction govern cutting in soft materials}
\begin{figure}[H]
\centering
\includegraphics[width=1\textwidth]{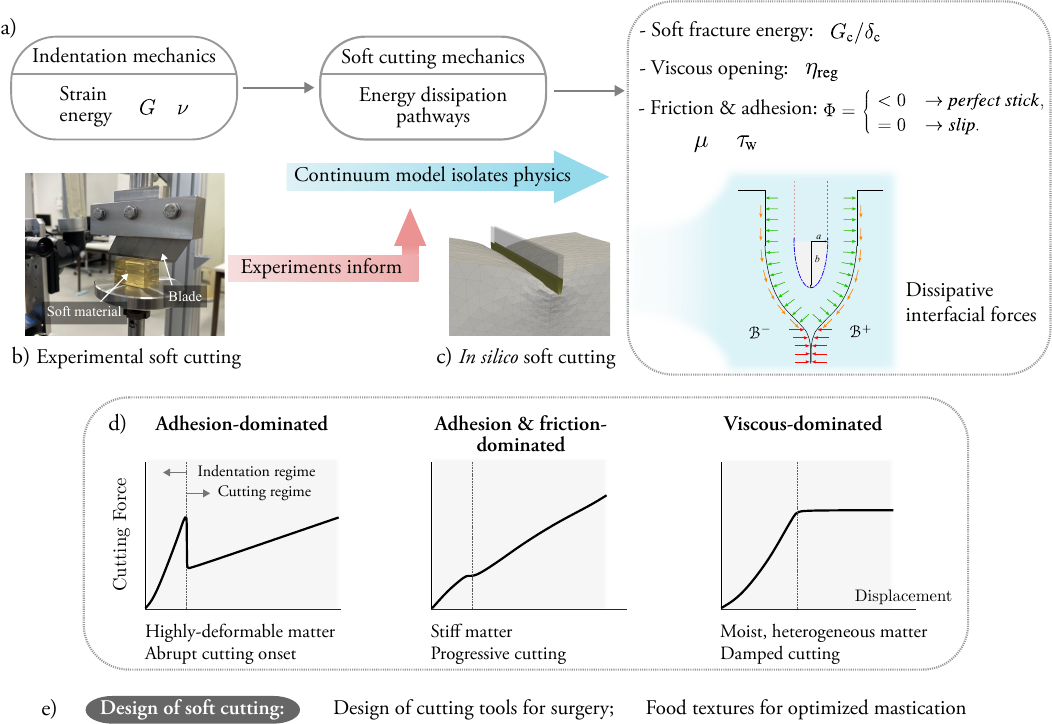}
\caption{\textbf{Overview of cutting mechanics in soft materials.} (a) Schematic illustration of the energy pathways in soft material cutting, including initial indentation, fracture propagation, and dissipation mechanisms. Strain energy stored during indentation is dissipated through the creation of new surfaces (fracture energy), viscous cohesive forces, and interfacial mechanisms such as friction and adhesion at the tool–material interface. The parameters of the unified model for soft cutting are direct descriptors of the cutting mechanisms, as described in Table~\ref{table:overview_parameters}. (b) Cutting experiments on soft materials reveal physical responses that inform (c) a coupled continuum model based on cohesive zone and contact mechanics, allowing the isolation of dominant mechanisms. The schematic highlights the balance of normal and tangential interfacial forces at the blade–material interface. (d) Representative force–displacement responses illustrate three major mechanisms in soft cutting mechanics: adhesion-dominated, adhesion–friction coupled, and viscous-dominated behavior. (e) Application to design of soft cutting for surgical tools and food texture engineering design for optimized mastication.}
\label{fig:setup_comprehensive}
\end{figure}

Soft materials exhibit complex mechanical responses when cut, involving a combination of bulk material deformation, fracture decohesion at the cutting interface, and frictional interactions with the cutting tool. As illustrated in Figure~\ref{fig:setup_comprehensive}a, the strain energy accumulated in the bulk material during the initial phase of deformation (indentation) is released and dissipated to drive the formation of new cutting surfaces and to overcome frictional resistance arising from contact with the cutting tool. Understanding how these mechanisms interplay is essential to predict and control cutting behavior, yet remains challenging due to their coupled, nonlinear nature.

\begin{figure}[h]
\centering
\includegraphics[width=1\textwidth]{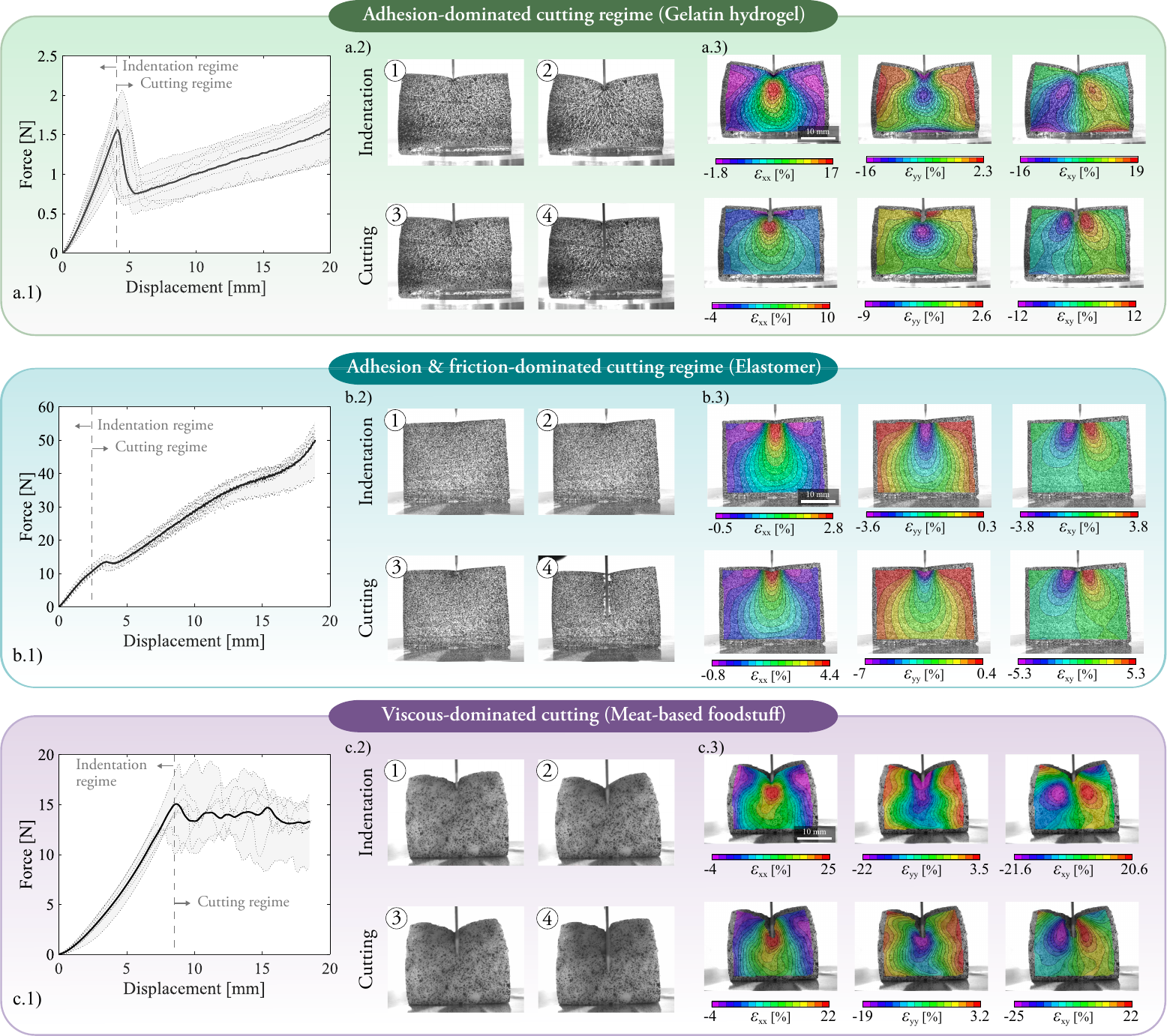}
\caption{\textbf{Experimental results for cutting tests.} Three cutting mechanisms---adhesion-dominated, adhesion- \& friction-dominated, and viscous dominated--- are exemplified with gelatin hydrogel, elastomer, and meat-based samples. (a-c.1) Cutting force versus indentation curves. Ten repetitions  are performed under the same test conditions. (a-c.2) Images of the surface of the samples during the indentation and the cutting regimes. The images correspond to the following values of displacement: for the hydrogel, \qty{2.70}{\milli \meter}, \qty{5.24}{\milli \meter}, \qty{6.18}{\milli \meter}, and \qty{14.14}{\milli \meter}; for the elastomer, \qty{0.62}{\milli \meter}, \qty{1.56}{\milli \meter}, \qty{2.29}{\milli \meter}, and \qty{14.14}{\milli \meter}; and for the meat-based foodstuf, \qty{6.24}{\milli \meter}, \qty{9.76}{\milli \meter}, \qty{10.40}{\milli \meter}, and \qty{14.14}{\milli \meter}. (a-c.3) Surface engineering strain fields---defined according to Section~\ref{sec:DIC_section}---for one of the experimental repetitions (one with a force--displacement curve close to the mean one is selected). 
Three stages are analyzed, i.e., two before cutting onset---the second one at the end of the initial indentation--- and one after the onset of cutting. The three stages correspond to the following values of displacement: for the hydrogel, \qty{2.7}{\milli \meter}, \qty{4.19}{\milli \meter}, and \qty{6.17}{\milli \meter}; for the elastomer, \qty{0.62}{\milli \meter}, \qty{1.56}{\milli \meter}, and \qty{2.29}{\milli \meter}; and for meat-based sample, \qty{4.80}{\milli \meter}, \qty{9.76}{\milli \meter}, and \qty{10.71}{\milli \meter}.
}
\label{fig:experim}
\end{figure}

We identify and describe three distinct cutting behaviors across the tested materials---gelatin hydrogel, elastomer, and meat-based foodstuff---that arise from the coupled effects of the material response at finite strains, the decohesion process along the cutting surface, and the frictional interaction between the material and the cutting tool. To that end, we leverage a custom-made experimental setup for soft cutting experiment (see Figure~\ref{fig:setup_comprehensive}b and Section~\ref{sec:cutting_experiments}). Additionally, we characterize the behavior of the materials under tensile deformation in Supplementary Figure\ref{fig:tensile}. The cutting force versus displacement curves in Figure~\ref{fig:experim} display different responses at the initial indentation regime \textit{ante} cutting onset and at the posterior cutting regime. 

The gelatin hydrogel exhibits an abrupt indentation-to-cutting transition characterized by a significant release of bulk strain energy, as illustrated in Figure~\ref{fig:experim}a. As a physically crosslinked polymer network, gelatin resists deformation through reversible physical bonds that stretch and accumulate energy as the blade compresses the material. The cutting force peaks early, around \qty{4}{\milli\meter} of displacement, then drops sharply, and subsequently increases linearly. At a critical stress level, the crosslinked network fails abruptly—particularly near the blade tip—leading to what we denote as a \emph{cohesive collapse}. This sharp drop in force reflects the sudden rupture of the material’s internal resistance. Interestingly, this behavior resembles that of materials with a hard outer shell or skin, where initial penetration is resisted until the blade breaks through, exposing a softer inner core. Following this rupture, the cutting force increases again, driven by frictional resistance as the blade interacts with the inner bulk. This friction is primarily adhesive in nature rather than due to Coulomb sliding, as will be demonstrated by the computational model in the subsequent sections.

The elastomer displays a smooth, nearly linear indentation-to-cutting transition, characterized by a gradual increase in force with displacement and no distinct fracture event, as shown in Figure~\ref{fig:experim}b. Unlike the gelatin hydrogel, where cutting is accompanied by a sudden release of stored energy, the elastomer resists cutting progressively. This behavior reflects its high toughness relative to the other materials: the blade advances through steady energy input, without abrupt failure or cohesive collapse. The absence of a sharp transition suggests that the material undergoes distributed deformation and energy dissipation, consistent with high tangential friction forces.

Lastly, the meat-based foodstuff, composed of finely ground meat and fat emulsified with binders such as starch, carrageenan, or soy protein, exhibits a smooth initial rise in cutting force, peaking around \qtyrange{8}{10}{\milli\meter} of blade displacement, followed by a fluctuating plateau at approximately \qtyrange{12}{16}{\newton} (see Figure~\ref{fig:experim}c). Unlike brittle or highly crosslinked materials, there is no substantial force drop following the peak; instead, the cutting force stabilizes, indicating a steady-state cutting regime. This behavior is characteristic of processed sausage products, which are typically wet and cohesive. The absence of a sharp drop in force at the onset of cutting reflects the absence of a brittle fracture mechanism. Instead, cutting initiates through ductile tearing and material flow, with no distinct transition between indentation and cutting—resulting in a gradual and continuous force response. This smooth transition, lacking a clear peak–drop–plateau sequence as observed in the hydrogel, highlights a cutting mechanism governed by viscous cohesive flow rather than crack propagation. Additionally, the fluctuations observed during the plateau phase suggest the presence of structural inhomogeneities within the material, which intermittently affect resistance as the blade progresses.

We illustrate the surface strain fields during cutting in Figure~\ref{fig:experim}a-c.3 and with supporting videos in Supplementary Videos 1, 2, and 3. Furthermore, we illustrate the three cutting mechanisms with cutting characterizations of additional foodstuffs, presented in Section \ref{sec:additional_foodstuff} and Figure~\ref{fig:additional_foodstuff}. The cutting responses of cheese, tofu, and marshmallow are discussed, highlighting the potential for engineering food textures optimized for mastication.

%\begin{fig\texttt{ure}[H]
%\centering
%\includegraphics[width=1\textwidth]{Fig_results_gelatine.pdf}
%\caption{\textbf{Results for experimental cutting tests on hydrogel 1.} Hydrogel 1 is made of a solution containing \qty{10}{\% w/v} of gelatin, with the liquid phase composed of \qty{50}{\% v/v} water and \qty{50}{\% v/v} glycerin. Before cutting onset, i.e., at the end of the initial indentation, and after cutting onset. (a) Force-indentation curves for 10 repetitions under the same test conditions. (b) Engineering strain fields (defined according to Section~\ref{sec:DIC_section}) for one of the experimental repetitions. The data corresponds to one of the experimental array for equal test conditions. Note that a test with a force--displacement curve close to the mean one is selected.}
%\label{fig:}
%\end{figure}}

%\begin{figure}[H]
%\centering
%\includegraphics[width=1\textwidth]{Fig_results_sylgard_v4.pdf}
%\caption{\textbf{Results for experimental cutting tests on Sylgard 184.} Test 4.}
%\label{fig:}
%\end{figure}

%\begin{figure}[H]
%\centering
%\includegraphics[width=1\textwidth]{Fig_results_sausage.pdf}
%\caption{\textbf{Results for experimental cutting tests on sausage.} Seven experimental repetitions.}
%\label{fig:}
%\end{figure}

\subsection{A coupled computational model disentangles physical contributions in soft cutting}
The analysis of the experimental results presented in the previous section faces limitations when testing specific hypotheses about the cutting mechanism as outlined in Figure~\ref{fig:setup_comprehensive}d. In particular, experimental techniques struggle to disentangle the contributions of tangential frictional stress---arising from contact pressure---from those of shear stress associated with wear. More broadly, the cohesive and frictional forces involved in cutting are intricately coupled and governed by the unknown constitutive behavior of the soft material. To address these challenges, we introduce a continuum-based computational model for the cutting of soft matter that replicates the experiments described in the previous sections (see the complete model in Section~\ref{sec:continuum_model}).

Serving as a virtual testbed, our continuum model enables the systematic isolation of individual physical contributions: the material's constitutive response (strain energy), the normal separation and the subsequent fracture-related energy dissipation at the cut surface due to contact with the cutting tool (see illustration in Figure~\ref{fig:setup_comprehensive}c and Figure~\ref{fig:setup}c for more detail), and the tangential friction dissipation resulting from relative motion, which combines effects from contact pressure, adhesive debonding, and wear, as outlined in Figure~\ref{fig:setup_comprehensive}a. The model enables direct simulation of indentation, fracture onset, and progressive cutting under realistic conditions. The model is modular in structure: it integrates cohesive debonding along the cutting surface, a viscous regularization of surface opening, and shear stresses arising from both friction and adhesion. These mechanisms are coupled within a unified framework, enabling systematic parametric studies in which specific contributions can be selectively activated or deactivated. This modularity is essential for understanding the complex physical interactions that govern the cutting process in soft materials.

\begin{figure}[H]
\centering
\includegraphics[width=0.93\textwidth]{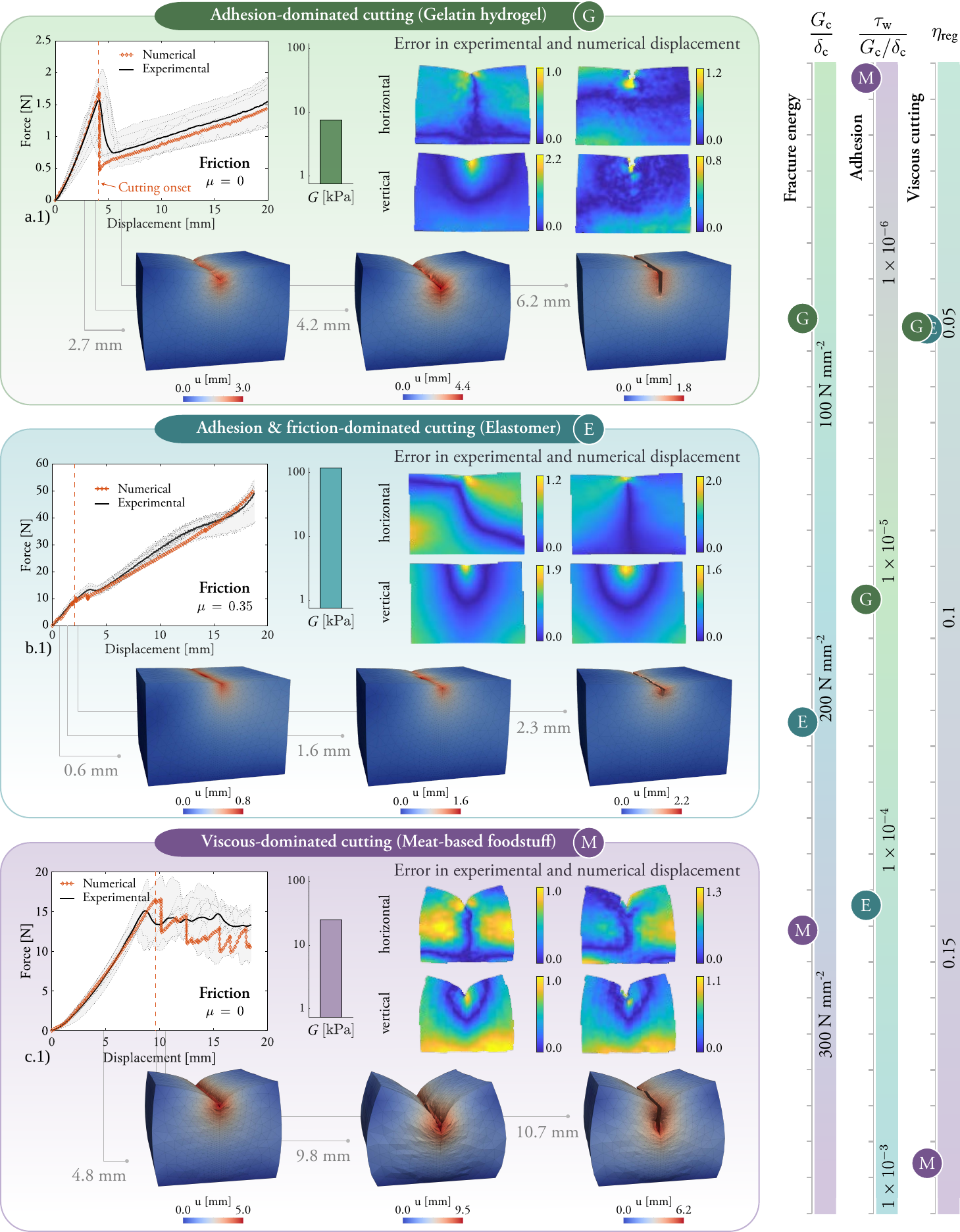}
\caption{\textbf{Virtual experiments obtained with the computational framework for cutting of soft materials.}  
(a-c.1) Numerical and experimental cutting force versus displacement curves for the gelatin hydrogel, elastomer, and meat-based foodstuff. 
(a-c.2) Stiffness of the sample as the parameter of the neo-Hookean model. 
(a-c.3) Relative error fields of the virtual experiments. The relative error field normalized by the average numerical displacement is computed as $|u_{x,\text{exp}}-u_{x,\text{num}}|/\left[N^{-1}\sum_1^N|u_{x,\text{num}}|\right]$ [-] and $|u_{y,\text{exp}}-u_{y,\text{num}}|/\left[N^{-1}\sum_1^N|u_{y,\text{num}}|\right]$ [-], and computed for two steps along the cutting process: one during indentation---before the onset of cutting--- and one in the cutting regime. 
(a-c.4) Frames of the deformed samples along indentation and cutting in the virtual experiments for the three materials. 
The cutting model is calibrated with the material parameters, fracture, friction, and viscous parameters in Table~\ref{table:overview_parameters}. All samples have the same cutting length (in the direction of the cutting tool) of \qty{30}{\milli \meter} and height of \qty{21}{\milli \meter}. The gelatin hydrogel and elastomer samples have a width of \qty{30}{\milli \meter} while meat-based samples have a width of \qty{25}{\milli \meter}. The error fields are shown for one representative of all experimental repetitions. 
}
\label{fig:model}
\end{figure}

As shown in Figure~\ref{fig:model}, the force-displacement curves from the virtual experiments accurately capture the distinct cutting responses observed in the physical tests, remaining within the bounds of experimental variability. For the gelatin hydrogel, the sharp drop in cutting force around \qty{4}{\milli\meter} of displacement reflects the sudden energy release associated with the indentation-to-cutting transition. In contrast, the elastomer exhibits a smooth transition with little to no reduction in cutting force, while the meat-based material shows a pronounced force plateau, consistent with experimental observations, following the onset of cutting. Moreover, the relative error between the experimentally measured and numerically simulated surface displacement fields confirms the overall strong agreement in deformation patterns between real and virtual experiments. The serrated shape of the force-displacement curve in the cutting regime is potentially related structural inhomogeneities in the material.

To interpret the experimental observations across the three distinct cutting behaviors, we rely on a reduced set of model parameters, each directly associated with a specific physical mechanism, as summarized in Table~\ref{table:overview_parameters}. After calibrating the model to reproduce the cutting response, we define three key dimensionless parameters, together with a viscosity parameter, that capture the underlying mechanics: $\frac{G_\text{c}/\delta_\text{c}}{G}$, the ratio of cohesive debonding energy to the material’s strain energy, governs the indentation-to-cutting transition and the steady-state cutting force; $\frac{\tau_\text{w}}{G_\text{c}/\delta_\text{c}}$ characterizes the contact shear stress from adhesion and wear relative to the cutting force; $\mu$ represents the coefficient of Coulomb friction; and $\eta_\text{reg}$ (in seconds) describes the rate-dependent viscous debonding of the cutting surface. 
The gelatin hydrogel is modeled with intermediate values $\frac{G_\text{c}/\delta_\text{c}}{G} = \qty{1.01e4}{}$ and $\frac{\tau_\text{w}}{G_\text{c}/\delta_\text{c}} = \qty{1.06e-5}{}$, no Coulomb friction ($\mu = 0$), and low viscosity $\eta_\text{reg} = \qty{0.05}{\second}$. 
The elastomer is described by a lower energy ratio $\frac{G_\text{c}/\delta_\text{c}}{G} = \qty{1.79e3}{}$, a higher adhesion ratio $\frac{\tau_\text{w}}{G_\text{c}/\delta_\text{c}} = \qty{1.89e-4}{}$, significant friction ($\mu = 0.35$), and low viscosity $\eta_\text{reg} = \qty{0.05}{\second}$. 
Lastly, the meat-based foodstuff is modeled with a high energy ratio $\frac{G_\text{c}/\delta_\text{c}}{G} = \qty{1.14e4}{}$, no adhesive shear stress ($\frac{\tau_\text{w}}{G_\text{c}/\delta_\text{c}} = 0$), no friction ($\mu = 0$), and a high viscous resistance $\eta_\text{reg} = \qty{0.2}{\second}$. 
The rulers at the right side in Figure~\ref{fig:model} provide a comparative overview of the parameters that govern the behavior of the cutting process.

\subsection{Material behavior, fracture, and friction modulate the stability of the indentation-to-cutting transition}

%Given the sharp tip of the cutting tool, the effects of friction and adhesion with the cutting tool can be neglected on the initial indentation phase. We verified that conducting additional experiments where the the surface in contact with the cutting tool during the initial indentation was coated with a lubricant oil. Inspecting the results no significant changes in the indentation-cutting transition where observed. As a consequence, 

Using the computational model, we investigate the critical transition from indentation to cutting, revealing conditions under which the transition is abrupt or smooth. The model captures the role of dissipative forces in stabilizing the cutting process and explains experimentally observed differences across material systems.

The transition from indentation to cutting involves overcoming both the cohesive resistance at the initiation of the interface and energy dissipation through frictional forces. Once this energy threshold is crossed, two primary responses may occur: (i) an abrupt drop in cutting force, driven by the rapid release of stored elastic energy when dissipative mechanisms are weak, or (ii) a gradual transition, enabled by sufficient energy dissipation through (a) frictional shear stresses at the tool–material interface and/or (b) viscous resistance within the cohesive zone. These behaviors are evident in the experimental results shown in Figure~\ref{fig:experim}, clarifying how different dissipation pathways control the onset and nature of cutting. The gelatin hydrogel exhibits an unstable force drop characteristic of low dissipation, while the elastomer and meat-based material show smoother transitions, reflecting dominant frictional and viscous damping, respectively.

For the gelatin hydrogel, the abrupt energy release and sharp drop in cutting force at the indentation-to-cutting transition are attributed to the low fracture energy $G_\text{c}$ relative to the material stiffness $G$, combined with moderate tangential adhesive stresses that permit a low cutting force immediately after decohesion initiates (Figure~\ref{fig:model}a.1). In contrast, the elastomer exhibits a smooth and stable transition, enabled by tangential dissipative forces from adhesion and wear, which absorb the energy released by the bulk during crack initiation (Figure~\ref{fig:model}b.1). Coulomb friction further stabilizes the process by resisting sudden force drops. This stabilizing effect is confirmed in Figure~\ref{fig:model_elastomer_mu0}, where setting the friction coefficient from 0.35 to zero leads to a more abrupt force decrease, highlighting the role of friction in smoothing the cutting transition. Finally, the meat-based foodstuff shows a delayed onset of cutting, consistent with a high \( G_\text{c} / G \) ratio (Figure~\ref{fig:model}c.1). In this case, large viscous cohesive forces at the cutting interface further resist decohesion, contributing to the postponed and smoother transition.

\begin{figure}[ht]
\centering
\includegraphics[width=1\textwidth]{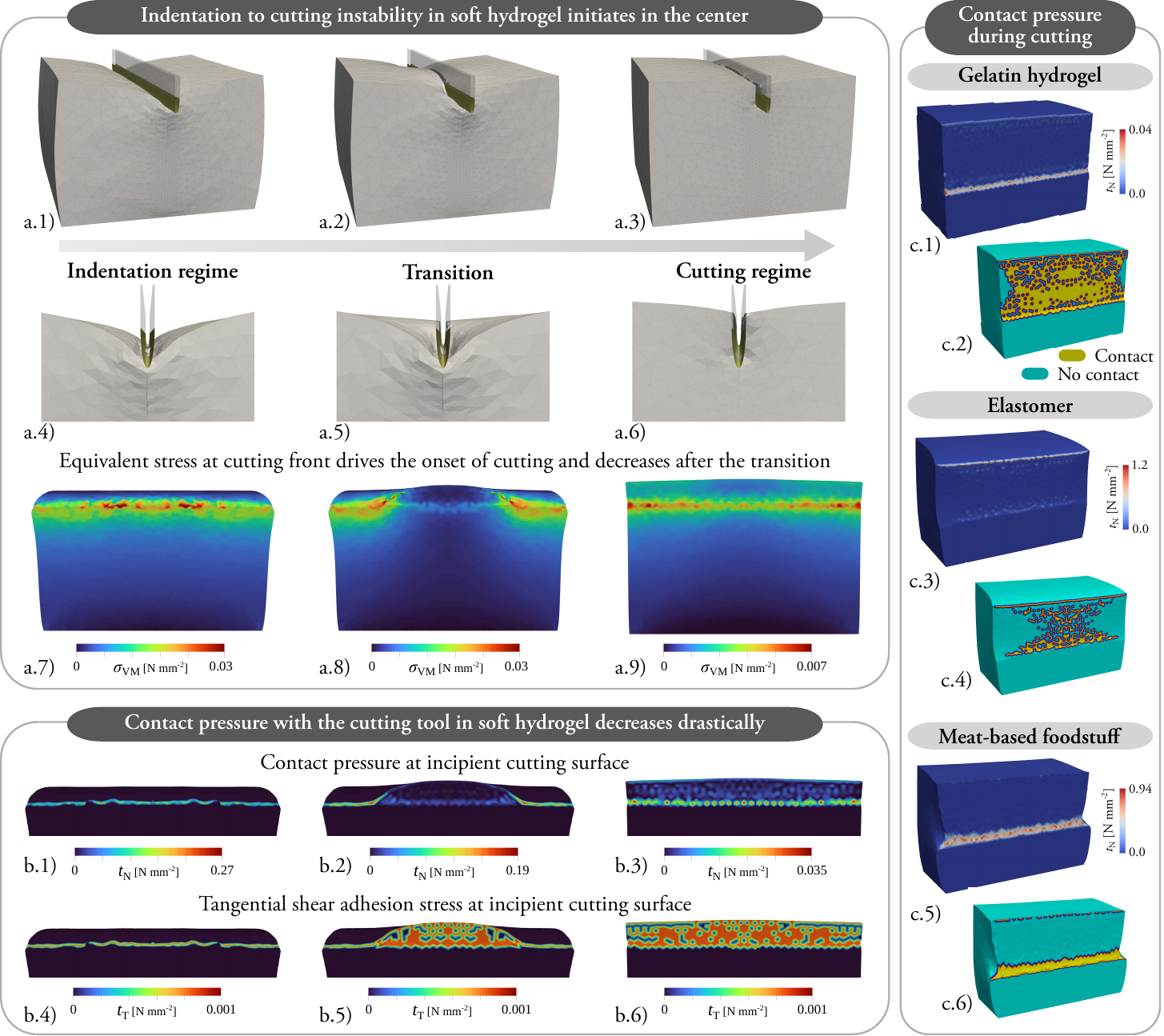}
\caption{\textbf{Constitutive behavior on the cutting surface investigated with the computational model: unstable indentation-to-cutting transition and contact pressure behind the crack front.} 
(a.1-3) Unstable indentation-to-cutting transition in a gelatin hydrogel at an indentation of the cutting tool where the strain energy of the continuum suffices to overcome the cohesive energy to create a new cutting surface. The onset of cutting occurs at the center of the sample beneath the center of the cutting tool. 
(a.4-6) Detail of the transition with an schematics of the cutting tool. 
(a.7-9) Equivalent Von Misses stress ($\sigma_\text{VM}$) at the cutting surface driving the indentation-to-cutting transition. 
(b.1-3) Decrease in the contact pressure $t_\text{N}$ in a gelatin hydrogel at the cutting surface. A normal stress prevents the interpenetration of material and the cutting tool along the transition. 
(b.4-6) Tangential contact stress $t_\text{T}$ due to adhesion and wear along the transition. 
(c.1,3,5) Contact pressure behind the crack front at a stage after the cutting tool has advanced in the cutting regime (a displacement of the cutting tool of \qty{13}{\milli \meter}) for the gelatin hydrogel, elastomer, and meat-based foodstuff. The contact pressure in the cutting surface is negligible behind the cutting front---at the sides of the blade. 
(c.2,4,6) Binary field that represents the state of normal contact. Coulomb friction is zero in the areas where contact is inactive and minimal when contact occurs but the contact pressure is negligible.}
\label{fig:indentation-to-cutting}
\end{figure}

The onset of cutting at sufficiently large indentations initiates at the center of the sample---directly beneath the center of the blade---and propagates laterally as an instability toward the blade edges. Figure~\ref{fig:indentation-to-cutting}a.1-6 illustrates this mechanism for the gelatin hydrogel, the material exhibiting the most pronounced cohesive collapse. In this central region, the bulk material deforms under near plane-strain conditions, as it is laterally confined by the surrounding material. This constraint limits deformation along the longitudinal direction of the blade, activating a stiffer volumetric response against the penetrating tool. As a result, the stress concentration in the central region is higher, leading to premature failure. The equivalent von Mises stress, $\sigma_\text{VM}$, along the indentation-to-cutting transition illustrates the mechanism. A peak, constant value of \qty{0.03}{\newton \per \milli \meter \squared} triggers the onset of cutting in the central region, as depicted in Figure~\ref{fig:indentation-to-cutting}a.7-9. Along the propagation of the cut in the longitudinal direction of the blade,  $\sigma_\text{VM}$ remains constant (Figure~\ref{fig:indentation-to-cutting}a.8) and it drops to a value of \qty{0.07}{\newton \per \milli \meter \squared} only at the end of the unstable transition corresponding to the beginning of the steady state cutting regime (see Figure~\ref{fig:indentation-to-cutting}a.9).

During the unstable indentation-to-cutting transition in the gelatin hydrogel, the contact pressure with the cutting tool drops sharply. Figure~\ref{fig:indentation-to-cutting}b illustrates the evolution of the pressure during the indentation-to-cutting transition: the maximum value of \qty{0.27}{\newton \per \milli \meter \squared} right \textit{ante} cutting onset drop progressively during the propagation of the cut to a steady value of \qty{0.035}{\newton \per \milli \meter \squared}. As a consequence, \textit{post} onset, the only remaining tangential contact stresses are due to adhesion and wear, and not Coulomb friction, as illustrated in Figure~\ref{fig:indentation-to-cutting}b.4-6, and Coulomb friction effectively vanishes. In the cutting regime, the resulting tangential contact force becomes proportional to the area of material–tool contact, with the stress limited by the maximum adhesive shear strength $\tau_\text{w}$. This value marks a transition from stick to slip conditions at the cutting interface.

\subsection{Adhesion, not Coulomb friction, controls tangential forces during cutting}

Contrary to conventional assumptions, our results show that adhesion and wear-related forces dominate tangential stresses during cutting, while Coulomb friction plays a negligible role due to low contact pressures. This finding reshapes the understanding of interfacial mechanics in soft cutting and guides improved material and tool design.

Not only immediately \textit{post} cutting onset, but also throughout the cutting regime, tangential contact stresses due to Coulomb friction vanish as the cutting tool advances into the material. The continuum model enables detailed exploration of the normal contact pressure distribution at advanced stages of the cutting process. As shown in Figure~\ref{fig:indentation-to-cutting}c.1, the gelatin hydrogel exhibits a peak contact pressure of \qty{0.04}{\newton\per\milli\meter\squared} localized at the cutting front. Behind the front, the pressure drops sharply, with only marginal, localized residual values persisting along parts of the cutting surface (Figure~\ref{fig:indentation-to-cutting}c.2). In the case of the elastomer, the highest contact pressure—reaching \qty{1.2}{\newton\per\milli\meter\squared}—appears along the upper edge of the cutting surface, while the regions trailing the crack front show minimal contact (Figure~\ref{fig:indentation-to-cutting}c.3,4). For the meat-based material, the peak contact pressure of \qty{0.94}{\newton\per\milli\meter\squared} is similarly located at the crack front, with near-zero pressure along the remaining surface due to a prevailing non-contact condition (Figure~\ref{fig:indentation-to-cutting}c.5,6). 

The continuum model for soft cutting reveals that friction does not contribute significantly to the tangential response observed experimentally, owing to the low contact pressures at the cutting surfaces. Instead, adhesive and wear-related contact forces fully account for the measured behavior. This interpretation is supported by the fact that the model is not sensitive to the value of the friction coefficient $\mu$ to reproduce the indentation-to-cutting transition of the gelatin hydrogel and meat-based materials, and by the observation of vanishing contact pressures during cutting. Even in the case of the elastomer, which has a non-zero $\mu$, the contact pressure along the cutting interface drops to negligible values. As shown in Figure~\ref{fig:model_elastomer_mu0}, simulations with and without friction yield almost identical slopes in the cutting force–displacement curves, indicating that the tangential resistance arises exclusively from adhesion and wear. This conclusion is further supported by the contact pressure profiles in Figure~\ref{fig:indentation-to-cutting}, which confirm that contact pressure along the cutting surfaces approaches zero during steady-state cutting.

Constraining the lateral expansion of the cube-shaped samples by restricting the motion of their side faces mimics the condition of a material embedded within a larger body, as would occur in alternative realistic cutting scenarios. This confinement may limit the ability of the material to deform freely in directions perpendicular to the cutting plane, thereby increasing the overall stiffness of the response against the penetrating tool. As a result, higher contact pressures may develop at the tool–material interface during cutting. We report additional cutting experiments on gelatin hydrogel samples with twice the original width, while maintaining the same cutting length, in Figure~\ref{fig:additional_BCs_exp}. Contrary to the previous hypothesis, the results demonstrate that increasing the lateral bulk volume does not significantly affect the cutting force response during the steady-state cutting regime, neither in terms of slope nor nominal force levels.

\section{Conclusion}
Our results reveal that the mechanics of cutting in soft materials cannot be captured by classical fracture or friction theories alone. Instead, the transition from indentation to cutting is governed by a coupled interplay of constitutive response, material dissipation, and adhesion. We showed that this transition can be either stable or unstable depending on material and interfacial parameters, that cutting initiation occurs centrally beneath the blade, and we described the role of adhesion in controlling the cutting force during the steady-state regime. Our model couples material and interfacial behavior on a comprehensive computational framework that substantiates, in the spirit of a virtual testbed, the understanding of adhesion-, friction-, and viscous-dominated cutting behaviors.

These findings provide a unifying framework for understanding why seemingly similar soft materials respond differently to the same cutting conditions. Beyond the specific systems studied here, the insights apply broadly to soft tissue manipulation, food processing, and the design of flexible materials and soft robotic interfaces. Future extensions may incorporate data-driven identification of material parameters from observed cutting behavior or time-dependent contact evolution. Our findings can therefore be adapted to other contexts such as surgical cutting of soft anisotropic materials \cite{Liu2021cutting} or food engineering \cite{Skamniotis2019}, and also to materials processing engineering in civil and manufacturing engineering contexts.

\newpage

\section{Experimental materials \& methods}\label{sec:experimental_methods}

\subsection{Materials}

\begin{itemize}[itemsep=0pt, parsep=0pt]
\item A gelatin hydrogel is produced mixing animal-based gelatin, water, and glycerin. 
The solution contains \qty{10}{\% w/v} of gelatin, with the liquid phase composed of \qty{50}{\% v/v} water and \qty{50}{\% v/v} glycerin. The solution was casted into a form and solidified at \qty{10}{\celsius}. The samples were manufactured by mixing for \qty{10}{\minute}, followed by storage in the fridge for \qty{1}{\hour}.
% \qty{50}{\gram} of gelatin powder, \qty{250}{\milli \liter} of water, \qty{250}{\milli \liter} of glycerin; and ii)  \qty{50}{\gram} of gelatin powder, \qty{250}{\milli \liter} of water, \qty{125}{\milli \liter} of glycerin.
\item Sylgard 184 (Dow Inc., Midland, Michigan, United States) was prepared mixing two raw phases in a 10:1---base to curing agent---volume mixing ratio. The crosslinked elastomer was cast into an open mold cured at \qty{90}{\celsius} during \qty{2}{\hour}. 
\item Processed meat-based foodstuff consisting of \qty{92}{\%} pork, complemented by drinking water and bacon. It contains---as disclaimed by the manufacturer---iodized salt (a mixture of table salt and potassium iodate), spice extracts such as fenugreek, chilli, ginger, cardamom, lovage, mace, and pepper, along with additional spices including chilli, coriander, onion, and marjoram. Dextrose is included as a sugar component, while diphosphates serve as stabilizers. Ascorbic acid acts as an antioxidant, and sodium nitrite is used as a preservative. The product is encased in pork intestine and finished with beechwood smoke.
\end{itemize}

The dimensions of the gelatin hydrogel and elastomer samples for cutting experiments are \qty{30}{\milli\meter} (width $w$) $\times$ \qty{30}{\milli\meter} (length in cutting direction) $\times$ \qty{21}{\milli\meter} (height $h$). For the meat-based samples, the width is reduced to $w = \qty{25}{\milli\meter}$ due to the difficulty of shaping larger specimens and the length in the direction of the blade (cutting length) is kept equal.

The dimensions of the gelatin hydrogel, elastomer, and meat-based samples for tensile tests experiments are \qty{12}{\milli \meter} (width $w$) $\times$ \qty{3}{\milli \meter} (thickness $t$) $\times$ \qty{30}{\milli \meter} (length $l$).

\subsection{Cutting experiments}\label{sec:cutting_experiments}
\begin{figure}[h]
\centering
\includegraphics[width=0.75\textwidth]{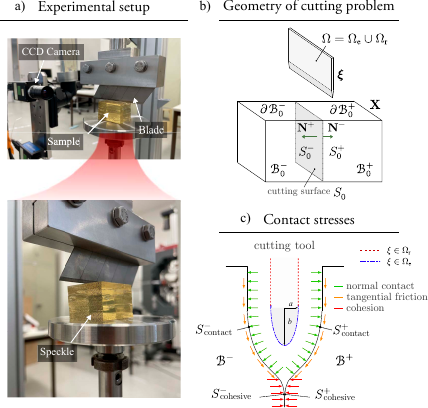}
\caption{\textbf{Overview of the experimental and virtual setup for cutting of soft materials.} 
(a) Experimental setup used for cutting experiments. 
(b) Schematics of the solid and cutting tool with nomenclature of the medium, its boundaries, and cohesive surface.
(c) Schematics of the balance of interfacial forces---contact-normal and tangential---forces on the contact boundary with the cutting tool, and cohesive forces on the crack front. The deformable medium is split into two regions through normal and tangential contact interactions with the cutting tool, which is modeled as a smooth mathematical surface. The tool comprises two segments: an upper region, $\Omega_\text{r}$, representing a rectangular section of infinite vertical extent, and a lower region, $\Omega_\text{e}$, forming an ellipsoidal tip. At the interface between the two halves of the medium, cohesive tractions are applied to resist separation, as governed by a cohesive law that depends explicitly on the displacement jump across the interface. }
\label{fig:setup}
\end{figure}

A universal tensile testing machine (Inspekt S 5 kN, Hegewald \& Peschke, Nossen, Germany) was adapted to perform cutting experiments using a \qty{0.7}{\milli\meter}-thick blade (wolfcraft, Kempenich, Germany), as shown in Figure~\ref{fig:setup}a. A custom-designed fixture was mounted on the upper crosshead to hold the blade and apply a quasi-static compressive displacement at a rate of \qty{0.021}{\milli\meter\per\second}, which renders an average strain rate of \qty{0.001}{\per\second}. The samples were positioned on a custom-designed support plate fixed to the lower crosshead of the testing machine. To ensure consistent cutting conditions, the blade was cleaned with lubricating oil after each experiment to remove any residual material or debris. All tests were conducted at room temperature, i.e., \qty{23}{\celsius}.

\subsection{Digital Image Correlation}\label{sec:DIC_section}
Pictures of the crack tip and surrounding area were taken during cutting experiments. A monochromatic CCD sensor (DCS 2.0, LIMESS Messtechnik \& Software GmbH, Germany) with a resolution of $1024 \times 768$ and a lens with focal range \qty{50}{\milli \meter} and aperture $2.8-16$ (2.8/50-0902 Xenoplan, Schneider Kreuznach, Bad Kreuznach, Germany) were used to capture the cracking pattern. Images were acquired at a rate of \qty{1.01}{\hertz}. The DIC postprocessing suite VIC-2D (Correlated Solutions Inc, Columbia, South Carolina) was used to compute the displacement and strain fields. A step size of \qty{7}{} pixel and a subset of \qty{27}{} pixel were used. In DIC, the step size is recommended to be lesser than one third of the subset size. The strain field computed is obtained as the Lagrange strain tensor, $\mathbf{E}=\frac{1}{2}\left[\mathbf{C}-\mathbf{I}\right]$, with $\mathbf{C}$ the right Cauchy-Green tensor and $\mathbf{I}$ the second order identity tensor.

The error fields between the experimental and numerical displacement fields shown in Figure~\ref{fig:model} are computed through a multi-step procedure. First, the numerical displacement field obtained from the FE simulation is interpolated onto a structured, equally spaced grid that matches the spatial resolution of the DIC experimental data. This ensures spatial correspondence between the two datasets over the lateral surface of the sample. The interpolated numerical data is then transformed into a matrix format consistent with the DIC output, where each entry corresponds to a pixel location (or subset, depending on the subset size) with appropriate scaling between millimeters and pixels. To enable direct comparison, the experimental displacement field is embedded into a matrix of the same dimensions as the numerical grid, using zero-padding along the margins. The error field is then calculated as the pointwise difference between the experimental and numerical displacement fields, and visualized in the deformed configuration, using the experimentally measured horizontal and vertical DIC displacements to represent the deformation.

\subsection{Tensile experiments}
An universal tensile machine (Inspekt S 5 kN, Hegewald \& Peschke, Nossen, Germany) was used to perform tensile tests on pre-cut rectangular samples---initial notch of \qty{2.4}{\milli \meter}---with a quasi-static tensile loading rate of \qty{0.03}{\milli \meter \per \second}, which renders an average strain rates of \qty{0.001}{\per \second}. The clamps on the machine were actuated with air pressure. The force-displacement data was stored during the deformation of the samples. All tests were conducted at room temperature, i.e., \qty{23}{\celsius}.

\newpage
\section{Computational model for cutting of soft solids: finite strains, fracture, and contact}\label{sec:continuum_model}
We propose a unified framework that combines a discontinuous spatial discretization at a cohesive interface---referred to as the \textit{cutting cohesive surface}---with a contact formulation with an ideally rigid cutting tool. The model incorporates tangential friction governed by Coulomb's law---proportional to the contact pressure up to the point of slipping---as well as adhesion, modeled as a constant stress activated upon contact.

Consequently, the cutting resistance experienced by the tool arises from four main contributions:
\begin{enumerate}[label=\roman*), itemsep=0pt, topsep=0pt]
    \item the fracture toughness or cohesive strength of the material,
    \item tangential frictional stress at the interface due to relative motion between the tool and the material, i.e., Coulomb friction,
    \item adhesive debonding stress, and
    \item shear stress associated with \textit{wear}---that is, surface damage occurring in the sample.
\end{enumerate}

\subsection{Kinematic framework}
Let us consider a solid undergoing inhomogeneous deformation induced by a displacement discontinuity, as illustrated in Figure~\ref{fig:setup}b. This strong discontinuity defines the cutting plane. Initially, the body is represented by a single connected domain, $\mathcal{B}_0$, which subsequently separates into two disjoint parts. In the material configuration, the cutting cohesive surface—i.e., the plane of separation—is denoted by $S_0$, and it partitions the solid into two distinct halves
\begin{equation}\label{eq:}
\mathcal{B}_0=\mathcal{B}_0^- \cup \mathcal{B}_0^+.
\end{equation}

%\begin{figure}[H]
%\centering
%\includegraphics[width=0.4\textwidth]{Fig_geometry.pdf}
%\caption{\textbf{Schematics of the solid and cutting tool with nomenclature of the medium, its boundaries, and cohesive surface.} }
%\label{fig:geometrical_considerations}
%\end{figure}

The two halves lie on the associated plus and minus sides of the cut surface
\begin{equation}\label{eq:}
S_0=S_0^- \cup S_0^+.
\end{equation}

Simultaneously, let the cutting surface consist of contact and cohesive parts, as illustrated in Figure~\ref{fig:setup}c,
\begin{equation}\label{eq:}
S_{0}^\pm = S_{\text{contact},0}^{\pm} \cup S_{\text{cohesive},0}^{\pm}.
\end{equation}

The deformation of the medium is formulated in a finite strain framework. The deformation $\boldsymbol \varphi \left(\mathbf{X}\right)$ maps the positions in the material configuration $\mathbf{X} \in \mathcal{B}_0$ to the positions in the spatial configuration $\mathbf{x} \in \mathcal{B}$ \textit{via}
\begin{equation}\label{eq:}
\mathbf{x}=\boldsymbol \varphi\left(\mathbf{X}\right)=\mathbf{u}\left(\mathbf{X}\right)+\mathbf{X}.
\end{equation}

Here, $\mathbf{u}$ denotes the displacement field. In the spatial configuration, the two cutting surfaces are denoted by $S^+$ and $S^-$, and the corresponding deformed regions of the body are $\mathcal{B}^+$ and $\mathcal{B}^-$. A surface traction will be defined later to enforce cohesion and to allow for eventual separation resulting from cutting the solid. This separation is characterized by the displacement jump across the cutting surface
\begin{equation}\label{eq:}
\llbracket \mathbf{u} \rrbracket = \mathbf{u}^+ - \mathbf{u}^-.
\end{equation}

Let us now define the material unit normal vector $\mathbf{N}$ as the outward normal to the boundary of the solid domain. Similarly, let $\mathbf{N}^\pm$ denote the unit normals on the positive and negative sides of the cohesive surface $S_0$, each pointing outward from the respective subdomains $\mathcal{B}_0^\pm$ as illustrated in Figure~\ref{fig:setup}b. In the spatial configuration, let $\mathbf{n}$ represent the unit normal to the outer boundary of the deformed body $\mathcal{B}$, and let $\mathbf{n}^\pm$ be the corresponding unit normals on either side of the discontinuous cutting interface $S$.

Consistently, the normal and tangential components of the displacement jump read
\begin{equation}\label{eq:}
\llbracket \mathbf{u} \rrbracket_\text{n} = \left[ \llbracket \mathbf{u} \rrbracket \cdot \bar{\mathbf{n}} \right] \bar{\mathbf{n}}
\quad \text{and} \quad
\llbracket \mathbf{u} \rrbracket_\text{t} = \left[ \mathbf{I} - \bar{\mathbf{n}} \otimes \bar{\mathbf{n}}\right] \mathbf{u},
\end{equation}
with the average spatial unit normal vector computed as $\bar{\mathbf{n}} = \left[\mathbf{n}^- - \mathbf{n}^+ \right] / ||\mathbf{n}^- - \mathbf{n}^+||$, for $\mathbf{n}^\pm =\left[ \mathbf{F}^{-\text{T}}\cdot\mathbf{N}\right]^\pm / ||\mathbf{F}^{-\text{T}}\cdot\mathbf{N}||^\pm$. The negative sign in front of $\mathbf{n}^+$ is necessary to reverse the orientation of the vector, as the face normal in each subdomain is defined to point outward from its respective cell. Without reversing the orientation of one of the vectors, their normal contributions would cancel out due to opposing directions.

The average deformation mapping for the deformed cutting surface $S$ further characterizes the deformation of the cohesive surface, allowing to identify a unique deformed configuration of $S$. It reads
\begin{equation}\label{eq:}
\{ \mathbf{u} \} = \frac{1}{2} \left[\mathbf{u}^+ + \mathbf{u}^-\right].
\end{equation}

The Dirichlet and Neumann boundaries of the solid are denoted, respectively, by $\partial_\text{D}$ and $\partial_\text{N}$
\begin{equation}\label{eq:}
\partial_\text{D} \mathcal{B}_0 = \partial_\text{D} \mathcal{B}_0^+ \, \cup \, \partial_\text{D} \mathcal{B}_0^- \quad \text{and} \quad
\partial_\text{N} \mathcal{B}_0 = \partial_\text{N} \mathcal{B}_0^+ \, \cup \, \partial_\text{N} \mathcal{B}_0^-.
\end{equation}

Note that $\partial_\text{D} \mathcal{B}_0$ and $\partial_\text{N} \mathcal{B}_0$ do not include the cutting surface $S_0$.

The deformation gradient is defined as 
\begin{equation}\label{eq:}
\mathbf{F}=\nabla_0\mathbf{u} + \mathbf{I}
\end{equation}
\noindent with $\mathbf{I}$ the second-order identity tensor and $\nabla_0$ the gradient operator in the material configuration. Following the multiplicative isochoric-volumetric decomposition into volumetric ($\mathbf{F}_\text{vol}$) and isochoric ($\overline{\mathbf{F}}$) parts,
\begin{equation}\label{eq:}
\mathbf{F}=\mathbf{F}_\text{vol}\cdot\overline{\mathbf{F}},
\end{equation}
\noindent with
\begin{equation}\label{eq:}
\mathbf{F}_\text{vol}=\left[ \det \mathbf{F}\right]^{1/3}\mathbf{I} \quad \text{and} \quad 
\overline{\mathbf{F}}=\left[\det\mathbf{F}\right]^{-1/3}\mathbf{F},
\end{equation}
\noindent and $\det \mathbf{F}$ the determinant of $\mathbf{F}$.

\subsection{Strong and weak forms}

The strong formulation of the problem reads
\begin{align}
\boldsymbol \nabla_0 \cdot \mathbf{P} + \mathbf{b}_0 = \mathbf{0}, \quad & \text{in} ~ \mathcal{B}_0^\pm,  \label{eq:strong_form1}\\
\mathbf{P} \cdot \mathbf{N} = \mathbf{t}^{\mathrm{p}/\mathrm{c}}, \quad & \text{on} ~ \partial_\text{N} \mathcal{B}_{0}^\pm, \label{eq:strong_form2}\\
\mathbf{u} = \mathbf{u}^\mathrm{p}, \quad & \text{on} ~ \partial_\text{D} \mathcal{B}_{0}^\pm, \label{eq:strong_form3}\\
 \left[\mathbf{P}\cdot \mathbf{N} \right]^\pm = \mathbf{t}^{\mathrm{c},\pm} , \quad & \text{on} ~ S_{\text{contact},0}^{\pm},  \label{eq:strong_form5}\\
\mathbf{P}^+ \cdot \mathbf{N}^+ + \mathbf{P}^- \cdot \mathbf{N}^- = \mathbf{0}, \quad & \text{on} ~ S_{\text{cohesive},0}^{\pm}, \label{eq:strong_form4}
\end{align}
\noindent with $\mathbf{t}^{\mathrm{p}/\mathrm{c}}$ a generic traction that is either prescribed (superscript ``$\mathrm{p}$'') or due to contact with the cutting tool (superscript ``$\mathrm{c}$'') and $\mathbf{u}^\mathrm{p}$ a prescribed displacement.

The weak formulation of the field equations in Equations~\ref{eq:strong_form1}-\ref{eq:strong_form4} is obtained by multiplying with a test function $\delta \mathbf{u}$ and integrating by parts,
\begin{align}\label{eq:weak_form0}
- \int_{\mathcal{B}_0} \mathbf{P} : \nabla_0 \delta\mathbf{u} \, \text{d}V 
+ \int_{\partial_\text{N}\mathcal{B}_{0}} \mathbf{t}^{\mathrm{p}/\mathrm{c}} \cdot \delta\mathbf{u} \, \text{d}A
+ \sum_\pm \int_{S_{\text{contact},0}^\pm} \mathbf{t}^\mathrm{c}\cdot \delta \mathbf{u}  \, \text{d}A &
 \nonumber \\
+ \int_{S_{\text{cohesive},0}^+} \mathbf{P}^+\cdot \mathbf{N}^+\cdot\delta\mathbf{u}^+\text{d}A
+ \int_{S_{\text{cohesive},0}^-} \mathbf{P}^-\cdot \mathbf{N}^-\cdot\delta\mathbf{u}^-\text{d}A &
= 0.
\end{align}

For simplicity, and given that the soft cutting model neglects body forces, the corresponding term has been omitted from Equation~\ref{eq:weak_form0}.

A further manipulation of the weak form can be done to express it in terms of the discontinuous operators---jump and average---across the cutting surface. Taking $\mathbf{N}^-$ as reference normal vector at the cutting interface, the traction on the positive side can be expressed as $\mathbf{P}^+ \cdot \mathbf{N}^+ = - \mathbf{P}^+ \cdot \mathbf{N}^-$ and the re-formulated weak form
\begin{align}\label{eq:}
- \int_{\mathcal{B}_0} \mathbf{P} : \nabla_0 \delta\mathbf{u} \, \text{d}V 
+ \int_{\partial_\text{N}\mathcal{B}_{0}} \mathbf{t}^{\mathrm{p}/\mathrm{c}} \cdot \delta\mathbf{u} \, \text{d}A
+  \sum_\pm \int_{S_{\text{contact},0}^\pm} \mathbf{t}^\mathrm{c}\cdot \delta \mathbf{u}  \, \text{d}A
 &
 \nonumber \\
- \int_{S_{\text{cohesive},0}} \llbracket \mathbf{P}\cdot\delta\mathbf{u}\rrbracket \cdot \mathbf{N}^-\text{d}A
&= 0.
\end{align}

The split of the jump of the product as $\llbracket \mathbf{P}\cdot\delta\mathbf{u}\rrbracket =  \{ \mathbf{P} \} \cdot \llbracket \delta\mathbf{u}\rrbracket + \llbracket \mathbf{P}\rrbracket \cdot \{\delta\mathbf{u}\}$ renders
\begin{align}\label{eq:weak_form11}
 \int_{\mathcal{B}_0} \mathbf{P} : \nabla_0 \delta\mathbf{u} \, \text{d}V 
+ \int_{S_{\text{cohesive},0}} \{ \mathbf{P} \} \cdot \mathbf{N}^- \cdot \llbracket \delta\mathbf{u}\rrbracket  \, \text{d}A
+ \int_{S_{\text{cohesive},0}} \llbracket \mathbf{P} \rrbracket \cdot \mathbf{N}^- \cdot \{ \delta\mathbf{u}\}  \, \text{d}A &
= \nonumber \\
 \int_{\partial_\text{N}\mathcal{B}_{0}} \mathbf{t}^{\mathrm{p}/\mathrm{c}} \cdot \delta\mathbf{u} \, \text{d}A
+  \sum_\pm \int_{S_{\text{contact},0}^\pm} \mathbf{t}^\mathrm{c}\cdot \delta \mathbf{u}  \, \text{d}A, &
\end{align}
\noindent where $\llbracket \delta \mathbf{u} \rrbracket = \delta \left[ \mathbf{u}^+-\mathbf{u}^- \right]$ denotes the jump operator across the cutting surface.

Eventually, the balance of tractions in the cohesive part of the cutting surface, $S_{\text{cohesive},0}$, (Equation~\ref{eq:strong_form4}) cancels the integral $\int_{S_{\text{cohesive},0}} \llbracket \mathbf{P} \rrbracket \cdot \mathbf{N}^+ \cdot \{ \delta\mathbf{u}\}  \, \text{d}A$ in Equation~\ref{eq:weak_form11}, yielding the final most convenient variant of the weak form
\begin{equation}\label{eq:weak_form1}
 \int_{\mathcal{B}_0} \mathbf{P} : \nabla_0 \delta\mathbf{u} \, \text{d}V 
+ \int_{S_{\text{cohesive},0}} \{ \mathbf{P} \} \cdot \mathbf{N}^- \cdot \llbracket \delta\mathbf{u}\rrbracket  \, \text{d}A
= 
 \int_{\partial_\text{N}\mathcal{B}_{0}} \mathbf{t}^{\mathrm{p}/\mathrm{c}} \cdot \delta\mathbf{u} \, \text{d}A
+  \sum_\pm \int_{S_{\text{contact},0}^\pm} \mathbf{t}^\mathrm{c}\cdot \delta \mathbf{u}  \, \text{d}A.
\end{equation}

Note that the Galerkin approximation of the trial and test functions consists of piecewise functions in the Sobolev subspace, specifically $\mathbf{u} \in \mathcal{H}^1$ and $\delta \mathbf{u} \in \mathcal{H}^1$. These functions are continuous within the subdomains $\mathcal{B}_0^+$ and $\mathcal{B}_0^-$, but exhibit a discontinuity across the cohesive surface $S_0$.\footnote{An alternative implementation could rely on a fully discontinuous Galerkin finite-dimensional approximation of the fields \cite{Radovitzky2011}, which enables crack propagation along any finite element interface. While this offers a more general and flexible framework, in the context of cutting, the crack path is, in principle, constrained to follow the cutting surface.}

The duality, or work-conjugacy, relations between stress and deformation measures are also evident in Equation~\ref{eq:weak_form1}. Similar to conventional solids, the Piola stress tensor $\mathbf{P}$ performs work on the deformation gradient $\mathbf{F}$ throughout the bulk of the material, while the traction $\{ \mathbf{P} \} \cdot \mathbf{N}^-$ contributes to the work by acting on the displacement jumps across the discontinuous cutting interface. The opening displacement $\llbracket \mathbf{u} \rrbracket$ serves as a measure of deformation, while the tractions act as the corresponding conjugate stress measure.

\subsection{Soft fracture: Cohesive zone model}\label{sec:model_CZM}
In the context of cohesive debonding of the interface, the term $\int_{S_{\text{cohesive},0}} \{ \mathbf{P} \} \cdot \mathbf{N}^- \cdot \llbracket \delta\mathbf{u}\rrbracket  \, \text{d}A$ in the weak form (Equation~\ref{eq:weak_form1}) represents the virtual work done by the cohesive traction force across the cutting interface and can be rewritten in terms of the cohesive traction  (c.f., e.g., \cite{Dugdale1960,Yuan2018}),
\begin{equation}\label{eq:weak_form_cohesive}
\int_{S_{\text{cohesive},0}} \{ \mathbf{P} \} \cdot \mathbf{N}^- \cdot \llbracket \delta \mathbf{u} \rrbracket \, \text{d}A
:=
\int_{S_{\text{cohesive},0}}  \mathbf{T} \left(\llbracket \mathbf{u}\rrbracket \right) \cdot \llbracket \delta \mathbf{u}\rrbracket \, \text{d}A.
\end{equation}

The damage process, which involves the formation of new crack surfaces, is modeled using virtual cohesive elements. This is achieved by progressively reducing the cohesive traction as the separation between the bulk elements in a potential crack extension zone increases. Failure occurs when the cohesive traction vanishes upon exceeding a critical separation threshold. A traction-separation law governs the constitutive behavior of decohesion, which can be defined in terms of the cohesive free energy density per unit undeformed area, $W_\text{s}$, over $S_0$ \cite{Needleman1990,Ortiz1999}, along with the cohesive law $\mathbf{T} \left(\llbracket \mathbf{u}\rrbracket \right)$,
\begin{equation}\label{eq:}
\mathbf{T} \left(\llbracket \mathbf{u}\rrbracket \right) = \frac{\partial W_\text{s}}{\partial \llbracket \mathbf{u} \rrbracket}.
\end{equation}

In turn, the cohesive law can be expressed in terms of the normal and tangential components---representing the resistance to normal opening and sliding, respectively---as
\begin{equation}\label{eq:traction_1}
\mathbf{T} = \frac{\partial W_\text{s}}{\partial \llbracket \mathbf{u} \rrbracket_\text{n}} \mathbf{n} + 
\frac{\partial W_\text{s}}{\partial \llbracket \mathbf{u} \rrbracket_\text{t}} \frac{\llbracket \mathbf{u} \rrbracket_\text{t}}{||\llbracket \mathbf{u} \rrbracket_\text{t}||}.
\end{equation}

To simplify the formulation of mixed-mode cohesive laws, the effective opening displacement \cite{Camacho1996} is defined as
\begin{equation}\label{eq:}
\delta\left(\llbracket \mathbf{u}\rrbracket \right) = 
\begin{cases}
    \sqrt{ \llbracket \mathbf{u}\rrbracket_\text{n} \cdot \llbracket \mathbf{u}\rrbracket_\text{n}
+
\gamma_\text{t} \llbracket \mathbf{u}\rrbracket_\text{t} \cdot \llbracket \mathbf{u}\rrbracket_\text{t}},  & \text{if }  \llbracket \mathbf{u}\rrbracket_\text{n} \geq 0, \\
    \sqrt{\gamma_\text{t} \llbracket \mathbf{u}\rrbracket_\text{t} \cdot \llbracket \mathbf{u}\rrbracket_\text{t}} , & \text{if } \llbracket \mathbf{u}\rrbracket_\text{n} < 0, 
\end{cases}
\end{equation}
\noindent where $\gamma_\text{t}$ is a parameter that weights the normal and sliding opening displacements. This definition allows to prevent damage when the positive and negative sides of the cutting surface interpenetrate, i.e., when $\llbracket \mathbf{u}\rrbracket_\text{n} < 0$.

Letting $W_\text{s}$ be a function of the effective opening, Equation~\ref{eq:traction_1} can then be reformulated as
\begin{equation}\label{eq:}
\mathbf{T} = \frac{1}{\delta}\frac{\partial W_\text{s}\left(\delta\right)}{\partial \delta} 
\left[ \llbracket \mathbf{u} \rrbracket_\text{n} + \gamma_\text{t}^2 \llbracket \mathbf{u} \rrbracket_\text{t} \rrbracket \right] ,
\end{equation}
\noindent where $t\left(\delta\right) = \partial W_\text{s}\left(\delta\right) / \partial \delta$ defines the effective traction.

Let us adopt a potential based on Smith and Ferrante's law\footnote{The Smith and Ferrante's law \cite{Rose1981} is one of the seminal simplified representations to calculate the cohesive energy in solids.} (cf. \cite{Ortiz1999}), 
\begin{equation}\label{eq:Smith_Ferrante}
W_\text{s} = e \sigma_\text{c}\delta_c \left[1-\left[1+\frac{\delta}{\delta_\text{c}}\right] e^{-\delta/\delta_\text{c}}\right],
\end{equation}
\noindent with $e \approx 2.71828$, $\sigma_\text{c}$ the maximum cohesive normal traction and $\delta_\text{c}$ a characteristic opening displacement. Note that the prefactor $e$ in Equation~\ref{eq:Smith_Ferrante} normalizes the potential so that the minimum energy is exactly $-\sigma_\text{c}\delta_c$ at $\delta=\delta_\text{c}$. It is a scaling factor introduced for convenience and consistency.

In addition, we introduce a viscous regularization of the cohesive free energy density allows to regularize the opening of the cutting surface at the transition from indentation to cutting regimes. The dependence on the pseudo-time is introduced with the term $\frac{1}{2}\eta_\text{reg} \llbracket \dot{ \mathbf{u}} \rrbracket\cdot\llbracket \dot{\mathbf{u}} \rrbracket$, where the time rate of the jump of the displacement field can be approximated as $\llbracket\dot{ \mathbf{u}} \rrbracket \approx \left[\llbracket \mathbf{u} \rrbracket_i - \llbracket \mathbf{u} \rrbracket_{i-1} \right]/\Delta t$, with $\Delta t$ the pseudo-time increment. The regularization term is then added up to the cohesive energy as $W = W_\text{s} + \frac{1}{2}\eta_\text{reg} \llbracket \dot{\mathbf{u}} \rrbracket\cdot\llbracket \dot{\mathbf{u}} \rrbracket$. This will produce additional cohesive forces proportional to $\llbracket \dot{\mathbf{u}} \rrbracket$ that can damp the abrupt opening of the cutting surface at the onset of the cutting regime\footnote{We note that viscous cohesive forces according to our approach may not vanish after complete decohesion if $\dot{\llbracket \mathbf{u} \rrbracket}\neq \mathbf{0}$. This does not occur in our computations as both flanks of the cutting surface remain separated exactly by the cutting tool and tangential motion is minimal. A more general formulation of viscous cohesive forces may imply degrading such forces with the damage state variable $d$.}.

As a consequence, the cohesive law reads
\begin{equation}\label{eq:traction_separation0}
\mathbf{T} \left(\llbracket \mathbf{u}\rrbracket \right)  = e \frac{\sigma_\text{c}}{\delta_c} \left[ \text{e}^{-\delta/\delta_\text{c}}
\left[ \llbracket \mathbf{u} \rrbracket_\text{n} + \gamma_\text{t}^2 \llbracket \mathbf{u} \rrbracket_\text{t} \rrbracket \right] + \eta_\text{reg} \llbracket \dot{\mathbf{u}}  \rrbracket\right].
\end{equation}

For convenience and to enforce damage irreversibility, let $d$ represent a damage state variable\footnote{Note that the choice of $d$ as internal variable is equivalent to setting the maximum attained effective opening displacement $\delta_\text{max}$ as internal variable.}
\begin{equation}\label{eq:damage_variable}
d = \max_{t\in[0,t]} \left( 1-e^{-\delta/\delta_\text{c}} \right).
\end{equation}

As a consequence of the state variable, an unloading event, which results in $\dot{\delta} < 0$, leads to a linear dependence between the cohesive traction and the effective opening.

Fracture criteria in cohesive zone modeling are based on an energy balance for crack nucleation and propagation. Crack initiation occurs when the energy available to separate the cohesive zone is sufficient. In non-linear elastic materials, a computation of the $J$-integral establishes the following link\footnote{Equation \ref{eq:cohesive_energy} denotes the limit of the cohesive free energy density $W_\text{s}$ as the opening displacement approaches infinity.} between the critical energy release rate $G_\text{c}$ for crack propagation and the cohesive law (see~\cite{Ortiz1999}),
\begin{equation}\label{eq:cohesive_energy}
G_\text{c} = \int_0^{\infty} T\left(\delta\right) \text{d}\delta
\qquad \text{with} \qquad
T\left(\delta\right) = e \frac{\sigma_\text{c}}{\delta_c} \text{e}^{-\delta/\delta_\text{c}}.
\end{equation}

For the case of Smith and Ferrante's law in Equation~\ref{eq:Smith_Ferrante}, the fracture toughness reads
\begin{equation}\label{eq:}
G_\text{c} =  e \sigma_\text{c} \delta_c 
\quad \rightarrow \quad
e \frac{\sigma_\text{c}}{\delta_\text{c}} = \frac{G_\text{c}}{\delta_\text{c}^2}.
\end{equation}

\begin{remark}
The energy release rate $G$ is equivalent to the $J$-integral and the cohesive energy $\Gamma_0$ in the absence of energy dissipation around the crack tip. This equivalence holds only at crack initiation. Once the crack propagates, elastic unloading becomes inevitable, and the energy consumed during cracking is less than the $J$-integral. As a result, it cannot be directly quantified from the global energy balance \cite{Yuan2019}.
\end{remark}

In the sequel, the cohesive law can be re-written in terms of the damage variable and fracture toughness,
\begin{equation}\label{eq:traction_separation}
\mathbf{T} \left(\llbracket \mathbf{u}\rrbracket \right) = 
\frac{G_\text{c}}{\delta_\text{c}^2}  \left[\left[1-d\right]  
\left[ \llbracket \mathbf{u} \rrbracket_\text{n} + \gamma_\text{t}^2 \llbracket \mathbf{u} \rrbracket_\text{t} \rrbracket \right] + \eta_\text{reg} \llbracket \dot{\mathbf{u}} \rrbracket \right],
\end{equation}
\noindent with $d$ the damage variable defined in Equation \ref{eq:damage_variable}.

The fracture parameters $G_\text{c}$ and $\delta_\text{c}$ are both constitutive parameters that must be calibrated simultaneously to ensure that the fracture toughness $\Gamma_0$ aligns with experimental observations. For the sake of convenience, we calibrate the ratio $G_\text{c}/\delta_\text{c}$ rather than both individual parameters. Since $\delta_\text{c}$ also appears in Equation~\ref{eq:damage_variable}, its value needs to be prescribed independently. Here, it is important to note that $\delta_\text{c}$ must be sufficiently small to guarantee complete decohesion between the two halves, $\mathcal{B}_0^\pm$, across the width of the cutting tool. For convenience, we assume that $\delta_\text{c} \ll a$, where $a$ will be defined in the following sections as half the width of the cutting tool. This ensures that the contact pressure between the solid and the tool is only due to the deformation of the bulk material and not to residual cohesive forces---not fully degraded---between both sides of the cohesive surface.

As a practical criterion, we set $\delta_\text{c} = 2a / 10$, i.e., ten times smaller than the width of the cutting tool, taking the value of \qty{0.07}{\milli \meter} in this work. This is sufficiently small to guarantee full decohesion for a separation equal to the width of the cutting tool. Section~\ref{sec:cohesive_law} illustrates the cohesive law as a function of the opening displacement. The reader will observe that $||\mathbf{T}||\left(\llbracket \mathbf{u} \rrbracket\right)$ for a normal opening of $\llbracket \mathbf{u} \rrbracket_\text{n} = 2a =  \qty{0.7}{\milli \meter}$, i.e., the width of the blade as defined later, is nearly zero.

\subsection{Contact with the cutting tool on the cutting surface}
The formulation for contact between the material and cutting tool at the cutting surface is incorporated as contact traction forces on the $+$ and $-$ sides of the contact part of the cutting surface, $S_{\text{contact},0}^\pm$. To this end, the related term is retrieved from the weak form (Equation~\ref{eq:weak_form1}) and reformulated in terms of normal contact traction forces to prevent interpenetration ($\mathbf{t}_{\text{N}}^\mathrm{c}$) and tangential contact traction forces for friction with the cutting tool\footnote{We consider contact and friction separate phenomena to be modeled independently of the cohesive law.} ($\mathbf{t}_\text{T}^\mathrm{c}$), i.e.,
\begin{align}\label{eq:weak_traction_forces_a}
&\sum_\pm \int_{S_{\text{contact},0}^\pm} \mathbf{t}^\mathrm{c}\cdot \delta \mathbf{u}  \, \text{d}A 
 = \nonumber \\
&\underbrace{\int_{S_{\text{contact},0}^+} \mathbf{t}_\text{N}^{\mathrm{c},+}\cdot \delta \mathbf{u}^+  \, \text{d}A +
\int_{S_{\text{contact},0}^-} \mathbf{t}_\text{N}^{\mathrm{c},-}\cdot \delta \mathbf{u}^-  \, \text{d}A}_\text{Normal contact: Obstacle problem} +
\underbrace{\int_{S_{\text{contact},0}^+} \mathbf{t}_\text{T}^{\mathrm{c},+}\cdot \delta \mathbf{u}^+  \, \text{d}A +
\int_{S_{\text{contact},0}^-} \mathbf{t}_\text{T}^{\mathrm{c},-}\cdot \delta \mathbf{u}^-  \, \text{d}A}_\text{Tangential contact: Coulomb friction \& adhesion}.
\end{align}

\subsection{Contact with the cutting tool on the external boundary during indentation}
Furthermore, contact during the initial indentation on the outer boundary is modeled with the traction force boundary integral in the weak form (Equation~\ref{eq:weak_form1}) according to
\begin{equation}\label{eq:weak_traction_forces_b}
 \int_{\partial_\text{N}\mathcal{B}_{0}} \mathbf{t}^{\mathrm{p}/\mathrm{c}} \cdot \delta\mathbf{u} \, \text{d}A = 
 \int_{\partial_\text{N}\mathcal{B}_{0}} \mathbf{t}^{\mathrm{p}} \cdot \delta\mathbf{u} \, \text{d}A
+ \underbrace{ \int_{\partial_\text{N}\mathcal{B}_{0}} \mathbf{t}^{\mathrm{c}}_{\text{N}} \cdot \delta\mathbf{u} \, \text{d}A}_\text{Normal contact: Obstacle problem}
+ \underbrace{\int_{\partial_\text{N}\mathcal{B}_{0}} \mathbf{t}^{\mathrm{c}}_{\text{T}} \cdot \delta\mathbf{u} \, \text{d}A.}_\text{Tangential contact: Coulomb friction \& adhesion}
\end{equation}

Note that the term including $\mathbf{t}^\mathrm{p}$ refers to any additionally prescribed traction force on the external boundary.

\subsection{Normal contact with cutting tool: Obstacle problem and augmented Lagrangian formulation}\label{sec:model_normal_contact}
A \textit{slave surface} is defined as the boundary of the deformable body, denoted by $\partial \mathcal{B}_0$, whereas the \textit{master surface} is modeled as a smooth two-dimensional surface $\Omega \subset \mathbb{R}^3$, parameterized by coordinates $\boldsymbol{\xi} \in \Omega$, representing an ideally rigid cutting tool.

The gap function---also referred to as distance or penetration function \cite{Wriggers2006,DeLorenzis2017}---is defined as
\begin{equation}
g_\text{N} = \min_{\mathbf{x} \subset \partial \mathcal{B}} || \mathbf{x} - \boldsymbol \xi ||.
\end{equation}

Subsequently, the following states of the contact area can be defined as
\[
\begin{array}{l l}
    g_\text{N} < 0 & \text{no contact}, \\
    g_\text{N} = 0 & \text{perfect contact}, \\
    g_\text{N} > 0 & \text{penetration}.
\end{array}
\]

The requirement that the deformable solid must not penetrate the cutting tool gives rise to a contact constraint, which is mathematically expressed through the Karush-Kuhn-Tucker (KKT) conditions
\begin{equation}\label{eq:KKT}
g_\text{N} \leq 0, \qquad p_\text{N} \geq 0, \qquad g_\text{N} \, p_\text{N} = 0,
\end{equation}
\noindent where $p_\text{N}$ denotes the contact pressure that arises as a reaction force associated with the constraint $g_\text{N} = 0$ in the case of active contact.

We define the \textit{master surface} representing the cutting tool, as described in Section~\ref{sec:experimental_methods}, to consist of an ellipsoidal tip and a constant-thickness rectangular segment of the blade (see Figure~\ref{fig:setup}c). Let $\Omega = \Omega_\text{e} \cup \Omega_\text{r}$ denote the surface of the cutting tool, where the subscript ``e'' refers to the ellipsoidal part and ``r'' to the rectangular upper segment. Accordingly, an \textit{ad hoc} definition for the gap function is given by
\begin{equation}\label{eq:gap_function}
g_\text{N} = 
\begin{cases}
    g_{\text{N,e}} , & \text{if } \min || \mathbf{x} - \boldsymbol \xi || \rightarrow \boldsymbol \xi \in \Omega_\text{e}, \\
    g_{\text{N,r}} , & \text{if } \min || \mathbf{x} - \boldsymbol \xi || \rightarrow \boldsymbol \xi \in \Omega_\text{r}. \\
\end{cases}
\end{equation}

The geometric definition of the ellipsoidal tip and rectangular segment can be expressed \textit{via} the following formalism:
\begin{align}\label{eq:}
g_{\text{N,ellip.}} &= 1 - \left[\mathbf{x} - \mathbf{c}\right]^\text{T} \cdot \mathbf{A}_1 \cdot \left[\mathbf{x} - \mathbf{c}\right], \\
g_{\text{N,rect.}} &= 1 -\left[\mathbf{x}-\mathbf{c}\right]^\text{T}\cdot\mathbf{A}_2 \cdot \left[\mathbf{x} - \mathbf{c}\right],
\end{align}
\noindent where $\mathbf{c} = \left[c_1, c_2, c_3\right]^\text{T}$ is the position vector of the center of the ellipse, and $\mathbf{A}_1$ is a symmetric matrix representing the shape and orientation of the ellipse, modeling the tip of the cutting blade. For an ellipse lying in the $xy$-plane, we have $\mathbf{A}_1 = \text{diag}\left(a^{-2}, b^{-2}, 0\right)$, where $a$ and $b$ are the lengths of the semi-axes of the ellipse. For the blade used in the experiments (see Section~\ref{sec:cutting_experiments}), we set $a = \qty{0.35}{\milli \meter}$. On the other hand, $\mathbf{A}_2$ is a matrix that represents the width and orientation of a rectangular upper band — of infinite vertical length — modeling the constant-thickness part of the cutting blade. Letting $a$ represent the small semi-axis of the ellipse (which corresponds to the thickness of the cutting tool), we define $\mathbf{A}_2 = \text{diag}\left(a^{-2}, 0, 0\right)$.
%\begin{figure}[H]
%\centering
%\includegraphics[width=0.6\textwidth]{Fig_contact.pdf}
%\caption{\textbf{Schematics of Contact—Normal and Tangential—Forces on the Contact Boundary with the Cutting Tool, and Cohesive Forces on the Crack Front.} The deformable medium is split into two regions through normal and tangential contact interactions with the cutting tool, which is modeled as a smooth mathematical surface. The tool comprises two segments: an upper region, $\Omega_\text{r}$, representing a rectangular section of infinite vertical extent, and a lower region, $\Omega_\text{e}$, forming an ellipsoidal tip. At the interface between the two halves of the medium, cohesive tractions are applied to resist separation, as governed by a cohesive law that depends explicitly on the displacement jump across the interface.}
%\label{fig:contact_considerations}
%\end{figure}

%\subsubsection{Augmented Lagrangian regularization of normal contact problem}

To implement the obstacle problem we employ an augmented Lagrangian framework. The framework regularizes the normal contact problem defined in Equation~\ref{eq:KKT} as proposed in the works of Simo \textit{et al}. \cite{Simo1992}. Consequently, the normal traction force $\mathbf{t}^{\mathrm{c}}_{\text{N}}$ in Equations~\ref{eq:weak_traction_forces_a} and \ref{eq:weak_traction_forces_b} can be expressed in terms of a Lagrange multiplier $\lambda_\text{N}$ as
\begin{equation}\label{eq:normal_traction_augmented}
\mathbf{t}^{\mathrm{c}}_{\text{N}} = - \left\langle \lambda_\text{N}^\pm + \frac{\epsilon_\text{N}}{h} g_\text{N}\left(\mathbf{u}^\pm \right) \right\rangle \mathbf{n}^\pm,
\end{equation}
\noindent with the spatial normal vector computed as $\mathbf{n} = \mathbf{F}^{-\text{T}}\cdot\mathbf{N} / ||\mathbf{F}^{-\text{T}}\cdot\mathbf{N}||$, $\epsilon_\text{N}=\qty{1}{}$ the penalty parameter for the augmentation scheme, $h$ the mesh size calculated as the diameter of the cell (elements in the FE mesh)\footnote{We regularize the penalty with the mesh size, since in the limit $h\rightarrow0$, a fixed penalty parameter would cause the penalty contribution in the weak form to vanish, rendering the constraint ineffective.}, and $\langle \bullet \rangle$ Macaulay brackets.

\begin{remark}
The obstacle problem constraint can be satisfied even if $\epsilon_\text{N}$ is undersized through repeated application of the augmentation procedure, preventing at the same time ill-conditioning of the resulting matrix problem.
\end{remark}

The multiplier in Equation~\ref{eq:normal_traction_augmented} is then augmented in successive iterations according to an Uzawa update scheme\footnote{The basic Uzawa iteration \cite{Uzawa1958} consists of two steps: i) minimize the weak form to find $\mathbf{u}$ and reduce the violation of the constraint (gap function), keeping the Lagrange multiplier $\lambda_\text{N}$ fixed; ii) update the multiplier to penalize the constraint violation, proportionally to how much the violation remains.}, i.e.,
\begin{equation}\label{eq:update:ALN}
\lambda_{\text{N},k+1}^{^\pm} = \left\langle \lambda_{\text{N},k}^{^\pm} + \frac{\epsilon_\text{N}}{h} g_\text{N}\left(\mathbf{u}^\pm \right)\right\rangle.
\end{equation}

After successive iterations, Equation~\ref{eq:update:ALN} drives the normal gap $g_\text{N}$ to zero, effectively enforcing the contact constraint and transferring the corresponding contact traction to the Lagrange multiplier $\lambda_\text{N}$.

\subsection{Tangential contact with cutting tool: Coulomb friction, adhesion and penalty regularization}\label{sec:model_friction}

%\subsubsection{Coulomb friction}
Tangential contact forces accounting for friction and adhesion are introduced through $\mathbf{t}^{\mathrm{c}}_{\text{T}}$ in Equations~\ref{eq:weak_traction_forces_a} and \ref{eq:weak_traction_forces_b}. We distinguish between two primary mechanisms that give rise to shear stress: friction and adhesion. Friction is modeled via Coulomb's law and leads to slipping when the tangential contact traction $\mathbf{t}^{\mathrm{c}}_\text{T}$ exceeds the product of a friction coefficient $\mu$ and the normal pressure. Adhesion, on the other hand, arises from the bonding strength between the material and the cutting tool, as well as from the resistance to wear (i.e., surface damage) of the material~\cite{Goda2024b}. The adhesive shear strength can reach a maximum value denoted by $\tau_\text{W}$.

Depending on the magnitude of the normal pressure, the contact forces exerted by the cutting tool may be dominated either by friction---under high contact pressures---or by adhesion and wear resistance---under low contact pressures. In practical terms, the parameter $\tau_\text{W}$ serves to characterize the adhesive shear strength and governs the frictional response when the normal contact pressure is small.

The Karush-Kuhn-Tucker conditions for Coulomb friction read
%\begin{equation}\label{eq:flow_potential}
%\Phi:=t_\text{T}  - \left[\mu t_\text{N} + \tau_\text{W}\tanh\left(\eta^{-1} t_\text{N}\right)\right],
%\end{equation}
\begin{equation}\label{eq:}
\dot{\mathbf{u}}-\dot{\mathbf{c}} = \rho \frac{\partial \Phi}{\partial \mathbf{t}^{\mathrm{c}}_\text{T}},
\end{equation}
\begin{equation}\label{eq:}
\rho \geq 0,
\end{equation}
\begin{equation}\label{eq:}
\rho \, \Phi = 0,
\end{equation}
\noindent with $\rho$ a Lagrange multiplier and $\mathbf{c}$ the position vector of the cutting tool\footnote{In our formulation the center of the cutting tool, $\mathbf{c}$, is defined as the center of an ideally ellipsoidal cutting tool can be used. Nonetheless, the reader may note that only the relative motion $\dot{\mathbf{c}}$ of the cutting tool in the cutting direction is required.}, hence $\dot{\mathbf{c}}$ is its rate of change and $\dot{\mathbf{u}}-\dot{\mathbf{c}}$ is the relative motion of the contact surface of the solid and the cutting tool along the successive iterations. Note that the relative motion $\dot{\mathbf{u}}-\dot{\mathbf{c}}$ is tangential upon friction due to the interpenetration contact condition in Equation~\ref{eq:KKT}.

The slip criterion, $\Phi$, is  defined as
\begin{equation}\label{eq:flow_potential}
\Phi:=t^{\mathrm{c}}_\text{T}  - \left[\mu t^{\mathrm{c}}_\text{N} + \tau_\text{W}\Theta\left(t^{\mathrm{c}}_{\text{N}}\right)\right] \leq 0,
\end{equation}
\noindent and it establishes whether \textit{slip} or \textit{perfect stick} occur according to
\begin{equation}\label{eq:slip_condition}
\Phi = 
\begin{cases}
< 0 & \rightarrow \textit{perfect stick} \text{ condition},\\
= 0 & \rightarrow \textit{slip} \text{ condition}.
\end{cases}
\end{equation}

Further, $\Theta\left(t^{\mathrm{c}}_{\text{N},n+1}\right)$ is an activation function to activate tangential contact due to adhesion and wear constant forces such that
\begin{equation}\label{eq:piecewise_adhesion}
\Theta\left(t_{\text{N}}\right) = 
\begin{cases}
1 & \text{if } t^{\mathrm{c}}_\text{N} > 0,\\
0 & \text{otherwise. }
\end{cases}
\end{equation}

\begin{remark}
In the absence of normal contact, i.e., when $t^{\mathrm{c}}_\text{N} = 0$, the contribution of $\tau_\text{W}$ in the friction potential $\Phi$ (Equation~\ref{eq:flow_potential})---which accounts for tangential dissipative forces due to adhesion and wear---is deactivated, resulting in $t^{\mathrm{c}}_\text{T} = 0$. Once normal contact is established, i.e., $t^{\mathrm{c}}_\text{N} > 0$, the switching function $\Theta\left(t^{\mathrm{c}}_{\text{N},n+1}\right)$ transitions from zero to one, thereby activating the contribution of $\tau_\text{W}$ and extending the flow potential $\Phi$ accordingly.
%The parameter $\eta$ in the hyperbolic tangent function controls the sharpness of the activation. Smaller values of $\eta$ approach to the ideal piece-wise activation with the initiation of normal contact but is detrimental to the numerical performance of the algorithm. In this work, $\eta=\qty{1e-4}{}$.
\end{remark}

%\subsubsection{Penalty regularization of frictional obstacle problem}

The simplest approach to enforce the friction constraint involves transforming the constrained minimization problem into an unconstrained one by applying a penalty regularization to the frictional obstacle problem:
\begin{equation}\label{eq:}
\mathbf{t}_{\text{T},n+1}^\text{c,trial}=\mathbf{t}^{\mathrm{c}}_{\text{T},n}+ \frac{\epsilon_\text{T}}{h}\left[\mathbf{u}_{n+1}-\mathbf{u}_{n} + \mathbf{c}_{n+1}-\mathbf{c}_{n} \right],
\end{equation}
%\begin{equation}\label{eq:}
%\Phi_{n+1}^\text{trial}=t_{\text{T},n+1}^\text{trial}  - \left[\mu t_{\text{N},n+1} + \tau_\text{W}\tanh\left(\eta^{-1} t_{\text{N},n+1}\right)\right],
%\end{equation}
\begin{equation}\label{eq:}
\Phi_{n+1}^\text{trial}=t_{\text{T},n+1}^\text{c,trial}  - \left[\mu t^{\mathrm{c}}_{\text{N},n+1} + \tau_\text{W}\Theta\left(t^{\mathrm{c}}_{\text{N},n+1}\right)\right],
\end{equation}
\noindent with $\epsilon_\text{T}=\qty{1e8}{}$ the penalty parameter for Coulomb friction.

The return mapping is completed with the following expression for the traction force to be inserted in Equations~\ref{eq:weak_traction_forces_a} and \ref{eq:weak_traction_forces_b}
\begin{equation}\label{eq:}
\mathbf{t}^{\mathrm{c}}_{\text{T},n+1}=\mathbf{t}_{\text{T},n+1}^\text{c,trial}-\Delta \zeta\frac{\mathbf{t}_{\text{T},n+1}^\text{c,trial}}{\|\mathbf{t}_{\text{T},n+1}^\text{c,trial} \|},
\end{equation}
\noindent with $\Delta \zeta$ the traction correction in the direction in the trial tangential force. The correction vanishes for \textit{perfect stick} condition and it actuates under the \textit{slip} condition according to
\begin{equation}\label{eq:gap_function}
\Delta \zeta = 
\begin{cases}
    0 , & \text{if } \Phi_{n+1}^\text{trial} \leq 0, \\
    \Phi_{n+1}^\text{trial} , & \text{if } \Phi_{n+1}^\text{trial} > 0. \\
\end{cases}
\end{equation}

\subsection{Bulk material behavior: hyperelastic constitutive model}
Only the definition of a constitutive model for the bulk material remains to be defined. Let the energy density per undeformed volume $\Psi$ be composed of isochoric and volumetric contributions, following the decoupled representation
\begin{equation}\label{eq:total_energy_density}
%\begin{split}
  \Psi(\mathbf{F}) = \Psi_\text{iso}(\overline{\mathbf{F}}) + \Psi_\text{vol}(\det \mathbf{F}).%\\[4pt]
% &= \Psi_\text{vol}(\mathbf{F})+ \Psi_\text{mec}^\text{Eq}(\mathbf{F}_\text{iso})+\Psi_\text{mec}^\text{NEq}(\mathbf{F}_\text{iso},\mathbf{F}^\upsilon_\text{iso}) ,
%\end{split}
\end{equation}

The isochoric contribution to the total energy density in Equation~\ref{eq:total_energy_density} is defined according to the neo-Hookean model as
%\begin{equation}\label{eq:Psi_isochoric}
%\Psi_\text{iso}\left(\overline{\mathbf{F}}\right)=
%C_{1}\left[\overline{I}_1-3\right]+C_{2}\left[\overline{I}_1-3\right]^2+C_{3}\left[\overline{I}_1-3\right]^3,
%\end{equation}
\begin{equation}\label{eq:Psi_isochoric}
\Psi_\text{iso}\left(\overline{\mathbf{F}}\right)=\frac{G}{2} \left[ \mathbf{I} : \left[\overline{\mathbf{F}}^\text{T}\cdot\overline{\mathbf{F}}\right] - 3\right],
\end{equation}
\noindent with the isochoric invariant $\overline{I}_1=\text{tr} \left(\overline{\mathbf{F}}^\text{T}\cdot\overline{\mathbf{F}}\right)$ and $G$ the shear modulus. The calibrated values of $G$ are detailed in \ref{table:overview_parameters}.

For the volumetric contribution, we use a relation dependent on the bulk modulus that is adequate to recover the nearly incompressible behavior of elastomers, i.e.,
\begin{equation}
\Psi_{\text{vol}} \left( J \right) =
\frac{\kappa}{2} \left[J-1\right]^2,
\quad 
\textrm{with}
\quad
\kappa=\frac{2G\left[1+\nu \right]}{3\left[1-2\nu\right]},
\end{equation}
\noindent for bulk modulus $\kappa$ with Poisson ratio $\nu$ set to 0.49 to recover nearly incompressibility\footnote{The constitutive response to volumetric deformations plays a crucial role in determining the cutting force (see, e.g., \cite{Goda2025}). Notably, even slight reductions in Poisson’s ratio within the nearly incompressible regime can lead to significant changes in both the onset of cutting and the magnitude of the cutting force. In this work we implement a compressible model with Poisson's ratio close to the incompressible limit and an object of future works may be the implementation of a mixed formulation for incompressibility, e.g., with Lagrange multipliers.}.

Lastly, the Piola stress tensor can be derived from the energy density as the addition of isochoric and volumetric contributions. The isochoric contribution reads
%\begin{align}\label{eq:}
%\mathbf{P}_\text{iso} = \frac{\partial \Psi_\text{iso} \left(\overline{\mathbf{F}}\right)}{\partial \mathbf{F}} =
%J^{-1/3}\mathbb{K}:\frac{\partial \Psi_\text{iso}\left(\overline{\mathbf{F}}\right)}{\partial \overline{\mathbf{F}}}
%=  J^{-1/3}\mathbb{K} : 
%\left[2C_{1}+4C_{2}\left[\overline{I}_1-3\right]+6C_{3}\left[\overline{I}_1-3\right]^2\right]\overline{\mathbf{F}},
%\end{align}
\begin{equation}\label{eq:}
\mathbf{P}_\text{iso} = \frac{\partial \Psi_\text{iso} \left(\overline{\mathbf{F}}\right)}{\partial \mathbf{F}} =
J^{-1/3}\mathbb{K}:\frac{\partial \Psi_\text{iso}\left(\overline{\mathbf{F}}\right)}{\partial \overline{\mathbf{F}}}
= G \, J^{-1/3}\mathbb{K} :  \overline{\textbf{F}}  ,
\end{equation}
\noindent with the fourth-order mixed-variant projection tensor $\mathbb{K}=\mathbb{I}-\frac{1}{3}\mathbf{F}^{-\text{T}}\otimes\mathbf{F}$, and the volumetric contribution
\begin{align}\label{eq:Pvol_v}
\mathbf{P}_\text{vol}
=\frac{\partial \Psi_\text{vol}\left(\mathbf{F}\right)}{\partial \mathbf{F}}=J\frac{\partial \Psi_\text{vol}\left(\mathbf{F}\right)}{\partial J}\mathbf{F}^{-\mathrm{T}}
=\kappa\left[J^2-J\right]\mathbf{F}^{-\text{T}}.
\end{align}

\subsection{Numerical Implementation}
The weak form in Equation~\ref{eq:weak_form1} is implemented in the latest version of the open-source finite element platform FEniCS \cite{Baratta2023, Scroggs2022, Alnaes2014}: FEniCSx v0.9.0. The Galerkin discretization of each of the two subdomains next to the cutting surface is done with linear Lagrange polynomial basis functions. The implementation of the mixed continuous Galerkin framework---continuous subdomains ($\mathcal{B}_0^\pm$) and discontinuous cohesive interface ($S_0$)---submeshes were utilized as described by Bleyer in \cite{Bleyer2024}. A FE mesh with tetrahedral elements is used to discretize the computation domain with finite-dimensional approximation of functions. The mesh for the gelatin hydrogel and elastomer samples (width of \qty{30}{\milli \meter}, cutting length of \qty{30}{\milli \meter}, and height of \qty{21}{\milli \meter}) has \qty{47801}{} elements and the mesh for the meat-based foodstuff (width of \qty{25}{\milli \meter}, cutting length of \qty{30}{\milli \meter}, and height of \qty{21}{\milli \meter}) has \qty{45456}{} elements. The displacement at the bottom surface in contact with the plate is fixed to zero in all directions via Dirichlet boundary conditions. This mimics the nearly-ideal adhesion of the samples in the experiments. 

The high non-linearity of the variational problem requires staggered numerical resolution strategies:
\begin{itemize}[itemsep=0pt, topsep=0pt]
\item The decohesion of the material ahead the cutting tool modeled by means of the traction-separation law in Section~\ref{sec:model_CZM} (Equation~\ref{eq:traction_separation}) is addressed with a fixed-point resolution strategy.  For a fixed time step, the displacement field $\mathbf{u}_i$ at a time step $i$ is solved for a damage field fixed to a previously known value $d_{i-1}$ \cite{Bleyer2024}. Then, the damage field is updated to $d_{i}$ using $\mathbf{u}_i$. The calculation is repeated iteratively until the error between iterations is under a tolerance.
\item The stepwise activation of friction due to adhesion and wear in Section~\ref{sec:model_friction} (Equation~\ref{eq:piecewise_adhesion}) is evaluated at a previously known value of the contact pressure, i.e., $\Theta\left(t_{\text{N},i-1} \right)$. Once the displacement field $\mathbf{u}_i$ is determined, $\Theta$ is updated to $\Theta\left(t_{\text{N},i} \right)$. The iteration loop is the same as the aforementioned one for the numerical update damage on the cohesive interface.
\item Further, to prevent spurious decohesion due to oscillations of $t_\text{N}$, which might be non-positive spontaneously on the sides of the cutting tool, $\Theta\left(t_{\text{N},n+1}\right)$ is kept active once contact between the cutting surface $S_0$ and the cutting tool has occurred.
\end{itemize}

The transition from the initial indentation regime to the cutting regime is a highly nonlinear process that necessitates sufficiently small time steps to ensure convergence. To balance accuracy and efficiency, we employ an adaptive load-stepping strategy that dynamically adjusts the time step based on the evolution of the damage variable. Specifically, when the increment in damage between successive steps exceeds a threshold of $\Delta d_{i\rightarrow i+1} = 0.1$, the time step is reduced by a factor of 10, down to a minimum of $\Delta t = 0.0001$. Conversely, if the damage grows more moderately, the time step is increased by a factor of 5, up to a maximum of $\Delta t = 0.2$. The total pseudo-time is set to $t_\text{end} = 21$, which matches the maximum indentation depth of the cutting tool. This ensures that the indentation process is controlled directly via the simulation pseudo-time, with both quantities defined to be equal.

The viscous regularization of Equation~\ref{eq:traction_separation} works hand-in-hand with the adaptive load-stepping solver. High values of the pseudo-time rate of the jump displacement $\llbracket\dot{ \mathbf{u}} \rrbracket$ lead to high damping cohesive forces. However, the reader may note that the pseudo-time play a role in the approximation $\llbracket \dot{\mathbf{u}}  \rrbracket\approx \left[\llbracket \mathbf{u} \rrbracket_i - \llbracket \mathbf{u} \rrbracket_{i-1} \right]/\Delta t$. Should the simulation halt the displacement of the cutting tool with $\Delta t \rightarrow 0$, and provided that the transition from indentation to cutting regimes is an instability, the regularization forces would tend to infinity. As a consequence, the minimum value of the pseudo-time step needs to remain greater than zero.

%Table~\ref{table:overview_parameters} summarizes the parameters of the model.
%\begin{table}[H]
%\centering
%\caption{\textbf{Overview of the parameters of the model.}}
%\begin{tabular}{l  l}
%\hline
%\textbf{Type} & \textbf{Parameter} \\
%\hline
%\multirow{2}{*}{Constitutive behavior of the bulk}  & $G$: shear modulus \\
%& $\nu$: Poisson's ratio\\
%\hline
%\multirow{3}{*}{Fracture debonding} & $G_\text{c}$: fracture toughness\\
%& $\delta_\text{c}$: characteristic opening displacement\\
%& $\eta_\text{reg}$: viscous regularization\\
%\hline
%\multirow{2}{*}{Tangential contact (friction)} & $\mu$: Friction coefficient \\
%& $\tau_\text{W}$: Critical shear stress for adhesion and wear\\
%\hline
%\end{tabular}
%\label{table:overview_parameters}
%\end{table}

\newpage
\subsection{Overview of parameters of the model: physical meaning}
The parameters of the model are direct descriptors of the physics in the cutting process. Table \ref{table:overview_parameters} describes them in relation to the cutting mechanism.
\begin{table}[H]
\centering
\caption{\textbf{Overview of the parameters of the model in relation to cutting mechanisms.}}
\small 
\begin{tabularx}{\linewidth}{p{5cm}  X}
\hline
\textbf{Parameter} & \textbf{Description in relation to cutting mechanisms} \\
\hline
Shear modulus ($G$)  & Defines the stiffness of the bulk material and the energy stored during indentation. Influences how much strain energy is available to drive the propagation of a cut. \\
\hline
Poisson's ratio ($\nu$) & Governs the material's lateral response under compression. Affects the bulk deformation around the blade during indentation and cutting. \\
\hline
Fracture toughness ($G_\text{c}$) & Total energy required to create new cutting surfaces via cohesive failure. Determines the resistance to cutting initiation and propagation. Higher values lead to tougher, more cut-resistant behavior. \\
\hline
Characteristic opening displacement ($\delta_\text{c}$) & Maximum separation at which the cohesive traction reduces by a factor of 0.37. Sets the opening width needed to degrade cohesive forces—needs to be enough smaller than the cutting blade thickness. \\
\hline
Viscous regularization ($\eta_\text{reg}$) & Introduces rate-dependence in the cohesive zone to improve numerical stability and mimic dissipative effects observed during steady cutting of viscoelastic materials like processed food. \\
\hline
Friction coefficient ($\mu$) & Coulomb-type friction opposing relative sliding between blade and material, proportional to contact pressure. Significant during indentation and tight contact phases. \\
\hline
Critical shear stress for adhesion and wear ($\tau_\text{W}$) & Pressure-independent shear resistance due to adhesion and surface stickiness. Important for capturing the increasing cutting force observed in cohesive and sticky materials in the cutting regime. \\
\hline
\end{tabularx}
\label{table:overview_parameters}
\end{table}

\newpage
\section*{Data and Software Availability}
\noindent The data generated during this study will be made available upon reasonable request.

\section*{Acknowledgments}
\noindent The authors acknowledge support from the European Research Council (ERC) under the Horizon Europe program (Grant-No. 101052785, project: SoftFrac). Funded by the European Union. Views and opinions expressed are however those of the authors only and do not necessarily reflect those of the European Union or the European Research Council Executive Agency. Neither the European Union nor the granting authority can be held responsible for them.
\begin{figure}[H]
\includegraphics[width=0.3\textwidth]{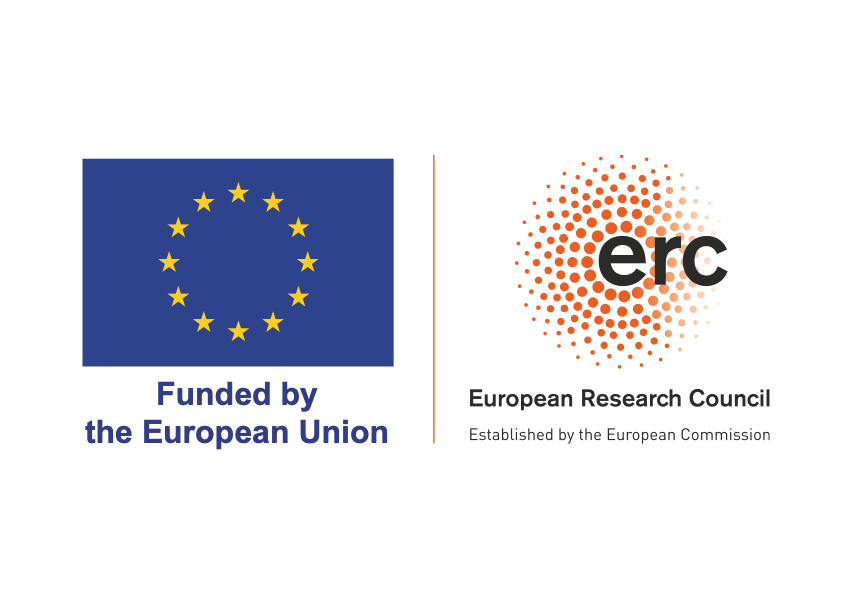}
\end{figure}

%\section*{Author Contributions}
%\noindent M.A.M.M. conceived the research, performed the experiments, developed the theory, implemented the computational model, and wrote the original manuscript. M.A.M.M. and P.S. conducted the formal analysis, discussion, and revised the paper.

\section*{Competing Interests}
\noindent The Authors declare no Competing Financial or Non-Financial Interests.

\newpage

\appendix

\section{Additional results for cutting of gelatin hydrogels with alternative compositions}
\begin{figure}[h]
\centering
\includegraphics[width=0.8\textwidth]{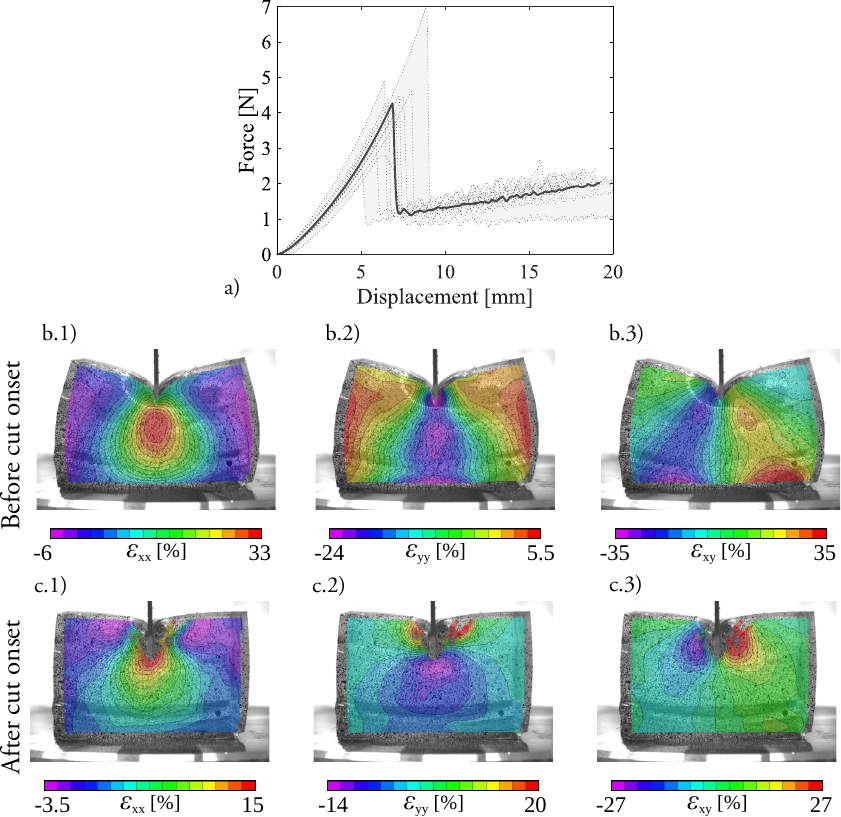}
\caption{\textbf{Results for experimental cutting tests on a first variant of the gelatin hydrogel.} A first variant is a solution containing \qty{13.33}{\% w/v} gelatin, with the liquid phase composed of \qty{66.67}{\% v/v} water and \qty{33.33}{\% v/v} glycerin. Before cutting onset, i.e., at the end of the initial indentation, and after cutting onset. (a) Force-indentation curves for ten repetitions under the same test conditions. (b) Engineering strain fields (defined according to Section~\ref{sec:DIC_section}) for one of the experimental repetitions. The data corresponds to one of the experimental array for equal test conditions. Note that a test with a force--displacement curve close to the average one is selected.}
\label{fig:addit_hydrogel_1}
\end{figure}

\begin{figure}[h]
\centering
\includegraphics[width=0.8\textwidth]{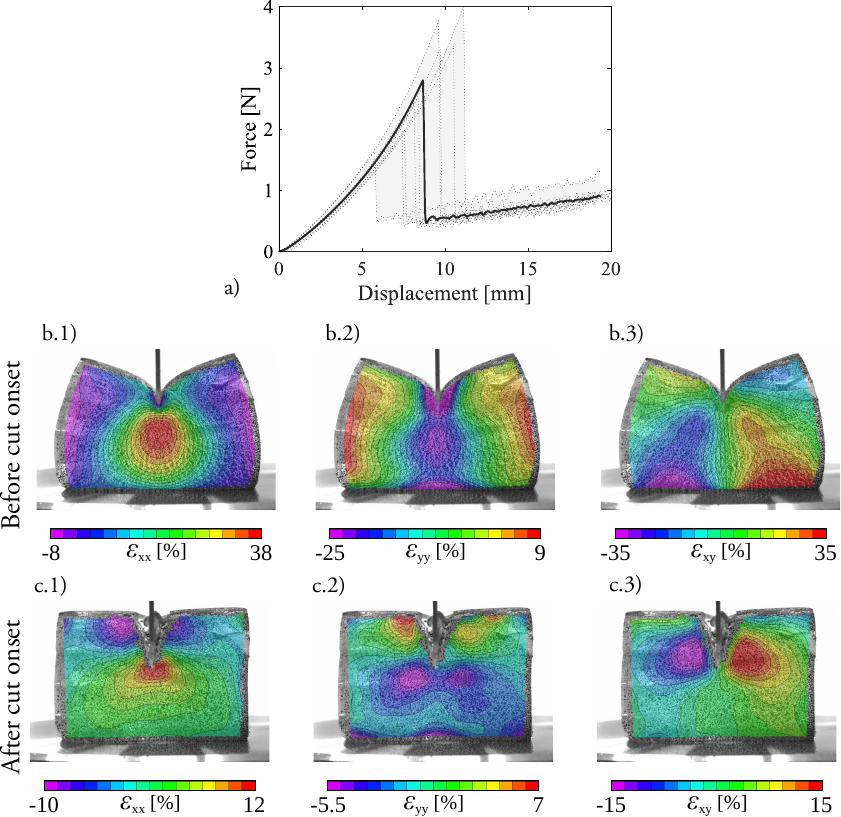}
\caption{\textbf{Results for experimental cutting tests on a second variant of the gelatin hydrogel.} A second variant is a solution containing \qty{10}{\% w/v} gelatin, with the liquid phase composed of \qty{100}{\% v/v} water. Before cutting onset, i.e., at the end of the initial indentation, and after cutting onset. (a) Force-indentation curves for ten repetitions under the same test conditions. (b) Engineering strain fields (defined according to Section~\ref{sec:DIC_section}) for one of the experimental repetitions. The data corresponds to one of the experimental array for equal test conditions. Note that a test with a force--displacement curve close to the average one is selected.}
\label{fig:addit_hydrogel_2}
\end{figure}

Two additional gelatin solutions were fabricated by mixing gelatin powder with a liquid mixture of water and glycerin in a different proportion: a solution containing \qty{13.33}{\% w/v} gelatin, with the liquid phase composed of \qty{66.67}{\% v/v} water and \qty{33.33}{\% v/v} glycerin; and a solution containing \qty{10}{\% w/v} gelatin, with the liquid phase composed of \qty{100}{\% v/v} water. The original one in the main manuscript is more hyperelastic than the two variants in this appendix, which are more brittle due to the lower content of glycerin. The results for the respective hydrogels are shown in Figures~\ref{fig:addit_hydrogel_1} and \ref{fig:addit_hydrogel_2}.

\section{Results for cutting of additional foodstuff}\label{sec:additional_foodstuff}
\begin{figure}[H]
\centering
\includegraphics[width=1\textwidth]{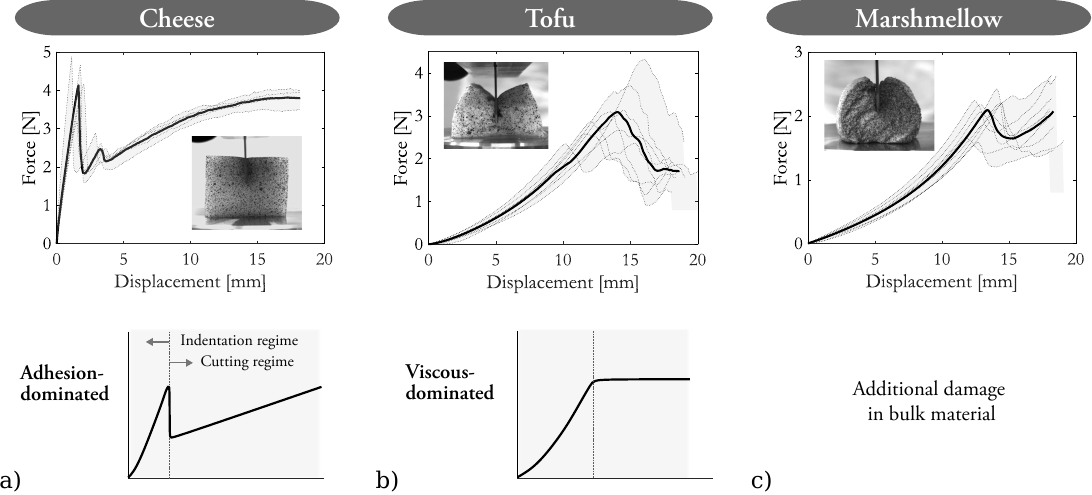}
\caption{\textbf{Results for cutting experiments on additional foodstuff.} Cutting force versus displacement curves for emmental cheese, tofu, and marshmellow. The dimensions for cheese samples are \qty{30}{\milli\meter} (width $w$) $\times$ \qty{30}{\milli\meter} (length in cutting direction) $\times$ \qty{21}{\milli\meter} (height $h$); for tofu samples, \qty{30}{\milli\meter} (width $w$) $\times$ \qty{30}{\milli\meter} (length in cutting direction) $\times$ \qty{21}{\milli\meter} (height $h$); for marshmellow samples,  \qty{18}{\milli\meter} (width $w$) $\times$ \qty{18}{\milli\meter} (length in cutting direction) $\times$ \qty{20}{\milli\meter} (height $h$). Insets: surface images of samples \textit{post} cutting onset.}
\label{fig:additional_foodstuff}
\end{figure}

Additional results for cutting experiments on foodstuffs: cheese, tofu, and marshmallow. The results are shown in Figure~\ref{fig:additional_foodstuff}. The foodstuffs are classified according to the three cutting mechanisms described in the main text: adhesion-dominated, adhesion- and friction-dominated, and viscous-dominated.

Emmental cheese exhibits an abrupt transition from indentation to cutting, followed by a second peak in the cutting force. Subsequently, the cutting force increases due to tangential adhesion with the cutting tool. This behavior may be influenced by the geometry of the cutting tool and the fat content of the cheese. Overall, the cutting response resembles that of the gelatin hydrogel discussed in the main text.

Tofu shows a smooth indentation-to-cutting transition, resembling the viscous cutting behavior observed in the meat-based foodstuff presented earlier. After the onset of cutting, the cutting force decreases progressively rather than abruptly. As previously argued for meat-based samples, this may be due to the heterogeneous structure of the material and its moisture content. Notably, the cutting force in tofu decreases over time, in contrast to the relatively stable force observed in the meat-based sample. This difference may point to the role of water diffusion in tofu, which could be explored in future studies.

Marshmallow displays a smooth indentation-to-cutting transition with noticeable remanent deformation in the bulk material. Although adhesion may occur, as suggested by the increasing cutting force after the onset of cutting, the cutting behavior does not clearly fit into any of the three proposed categories. Future work may investigate the constitutive behavior of this sugar-based material and its underlying damage mechanisms.

\newpage 

\section{Tensile characterization of the gelatin hydrogel, elastomer, and meat-based foodstuff}
\begin{figure}[h]
\centering
\includegraphics[width=1\textwidth]{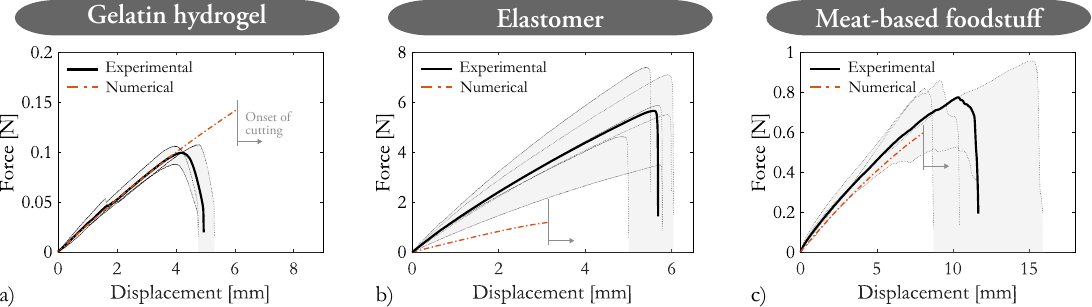}
\caption{\textbf{Results for tensile tests on pristine and pre-cut samples and validation of the computational model.} Experiments on gelatin hydrogel, elastomer, and meat-based foodstuff. The dimensions of the samples are \qty{12}{\milli \meter} (width) $\times$ \qty{30}{\milli \meter} (distance between clamps) $\times$ \qty{3}{\milli \meter} (thickness). The samples feature an initial notch of \qty{2.4}{\milli \meter} in the middle of a lateral of the sample. Three experimental repetitions are performed for the same test conditions. }
\label{fig:tensile}
\end{figure}

Characterization of the three materials investigated in the main manuscript via uniaxial tensile tests. Figure~\ref{fig:tensile} contains the experimental results and the cutting force versus displacement curves predicted by the computational model under virtual tensile loading conditions. Note that only the pre-cutting portions of the curves are shown, as the viscous regularization used in the model nonphysically dampens the debonding response even beyond the point of full damage.

\section{Calibration of parameters}\label{sec:calibration_parameters}
The parameters of the computational model, used in the main text to disentangle the physics in soft cutting, are calibrated to reproduce the experimental cutting curves and summarized in Table~\ref{table:overview_parameters}. 
\begin{table}[h]
\centering
\caption{\textbf{Calibration of parameters of the \textit{in silico} model.} The parameters of the unified model for soft cutting are direct descriptors of the cutting mechanisms, namely the bulk, cohesive fracture, and interfacial mechanisms.}
\begin{tabular}{l|ll|ll|ll|l}
\hline
\multirow{2}{*}{Material} & \multicolumn{2}{c|}{Bulk material} & \multicolumn{2}{c|}{Cohesive fracture} & \multicolumn{2}{c|}{Friction \& adhesion} &  Viscous\\
 & $G$ [MPa] &  $\nu$ [-] & $\delta_\text{c}$ [mm] & $G_\text{c}/\delta_\text{c}$ [\qty{}{\newton \per \meter \squared}] & $\mu$ [-] & $\tau_\text{W}$ [\qty{}{\newton \per \meter \squared}] & $\eta_\text{reg}$ [\qty{}{\second}] \\
\hline
Gelatin hydrogel & \qty{7.5e-3}{} & \qty{0.49}{} & \qty{0.07}{} & \qty{75.7}{} & \qty{0}{} & \qty{0.0008}{} & \qty{0.05}{} \\
Elastomer & \qty{120e-3}{} &  \qty{0.49}{} & \qty{0.07}{} & \qty{214.3}{} & \qty{0.35}{} & \qty{0.04}{} & \qty{0.05}{} \\
Meat-based foodstuff & \qty{25e-3}{} &  \qty{0.49}{} & \qty{0.07}{} & \qty{285.7}{} & \qty{0}{} & \qty{0}{} & \qty{0.2}{}  \\
\hline
\end{tabular}
\label{table:overview_calibrated_parameters}
\end{table}

We note that, for the elastomer (Sylgard 184), the calibration predicts a shear modulus about half the one observed in tensile experiments, as indicated in Figure \ref{fig:tensile}b. The response in stiffness observed in the tensile experiments agrees with the constitutive behavior determined in a biaxial characterization of the elastomer reported by the authors \cite{Moreno-Mateos2025}.  Overall, the differences may be due to slightly different constitutive behavior in tensile deformation and compression loading conditions, even related to volumetric deformation.

\section{Traction-separation law in the cutting surface}\label{sec:cohesive_law}
\begin{figure}[H]
\centering
\includegraphics[width=1\textwidth]{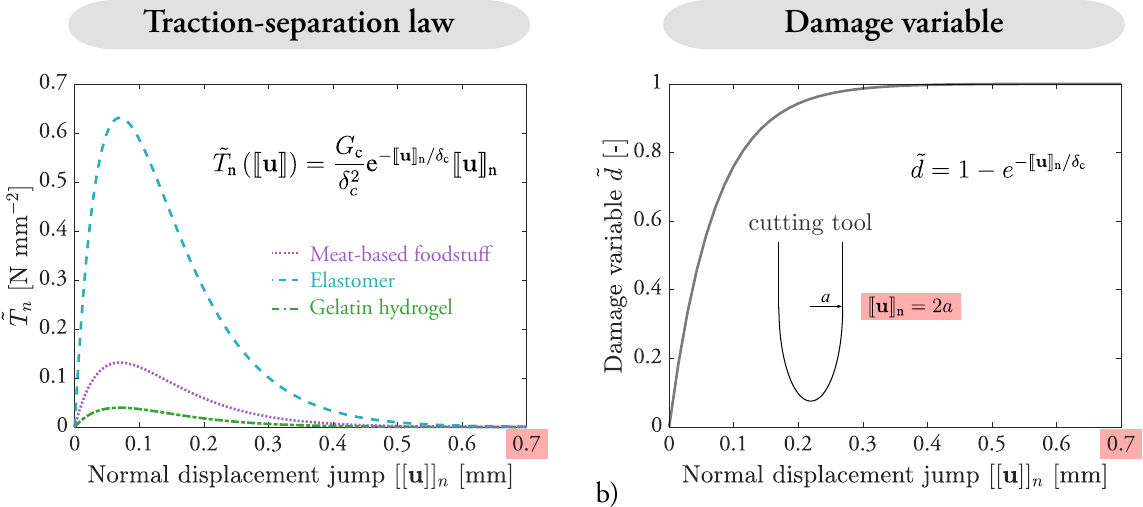}
\caption{\textbf{Traction-separation law on the cutting surface.} (a) Traction force per unit area as a function of the normal opening displacement across the cutting surface for the three materials: gelatin hydrogel, elastomer, and meat-based foodstuff. (b) Corresponding evolution of the damage variable with respect to the normal opening displacement. When the opening reaches the width of the cutting tool ($\llbracket \mathbf{u} \rrbracket_\text{n} = 2a$, see figure inset), the cohesive traction has fully vanished, and the damage variable approaches unity, indicating complete decohesion. For simplicity, the traction–separation law is plotted directly as a function of the normal opening displacement $\llbracket \mathbf{u} \rrbracket_\text{n}$ rather than the internal damage variable $d$ (cf. Equation~\ref{eq:traction_separation0} in the main text).}
\label{fig:tractionseparation_law}
\end{figure}

%\begin{equation}\label{eq:}
%\tilde{T}_\text{n} \left(\llbracket \mathbf{u}\rrbracket \right)  =  \frac{G_\text{c}}{\delta_c^2} \text{e}^{-\llbracket \mathbf{u}\rrbracket_\text{n}/\delta_\text{c}}
%\llbracket \mathbf{u} \rrbracket_\text{n} = 2a
%\end{equation}
%\begin{equation}\label{eq:}
% \tilde{d} = 1-e^{-\llbracket \mathbf{u} \rrbracket_\text{n}/\delta_\text{c}} 
%\end{equation}
The cohesive traction forces on the cohesive surface are modeled using traction–separation laws that ensure complete degradation of the tractions when the material undergoes a separation equal to the width of the cutting tool. This behavior is enforced through the inclusion of an exponential term in the law and by selecting a sufficiently small characteristic opening displacement (see Table~\ref{table:overview_calibrated_parameters} for its value). A plot of the damage variable $d$ as a function of the normal opening displacement of the cohesive surface confirms that it approaches 1 (i.e., full degradation) for an opening $\llbracket \mathbf{u} \rrbracket_\text{n} = 2a$, with $a$ half the thickness of the cutting tool. This is illustrated in Figure~\ref{fig:tractionseparation_law}.

\newpage

\section{Additional cutting simulation for the elastomer material without Coulomb friction ($\mu=0$)}
\begin{figure}[h]
\centering
\includegraphics[width=0.45\textwidth]{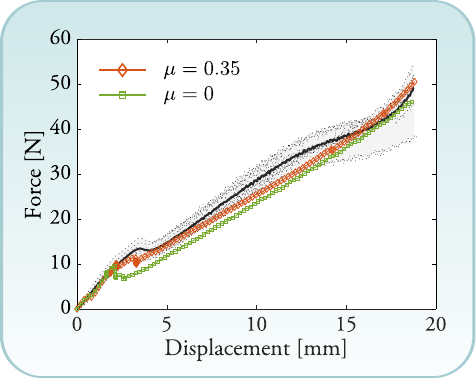}
\caption{\textbf{Additional simulation for the elastomer material to assess the effect of Coulomb friction at the indentation-to-cutting transition.} To isolate its influence, the friction coefficient is set to $\mu = 0$ and the resulting cutting force–displacement curve is compared against the reference case with active Coulomb friction ($\mu = 0.35$). This comparison reveals the impact of interfacial shear resistance on the force response during the onset of cutting, which smooths the transition.}
\label{fig:model_elastomer_mu0}
\end{figure}

A simulation with a friction coefficient $\mu=0$ yields a cutting force, after the onset of cutting, with a slope nearly identical to that obtained when Coulomb friction is included, as shown in Figure~\ref{fig:model_elastomer_mu0}. This suggests that the tangential resistance arises primarily from adhesion and wear, rather than from Coulomb friction.

\newpage
\section{Additional cutting experiments on gelatin hydrogel samples to investigate the effect of boundary conditions}
To investigate the influence of sample geometry on the contact pressure along the cutting surfaces, additional experiments were performed on samples with double the width (perpendicular to the blade) compared to the original ones. The evolution of the cutting force with the displacement of the cutting tool is compared to that of the narrower samples. The results show that increasing the amount of bulk material lateral to the cutting tool has a negligible effect on the cutting force during the cutting regime. This indicates that the interfacial mechanics remain largely unchanged, with negligible contact pressure and Coulomb friction.
\begin{figure}[H]
\centering
\includegraphics[width=0.65\textwidth]{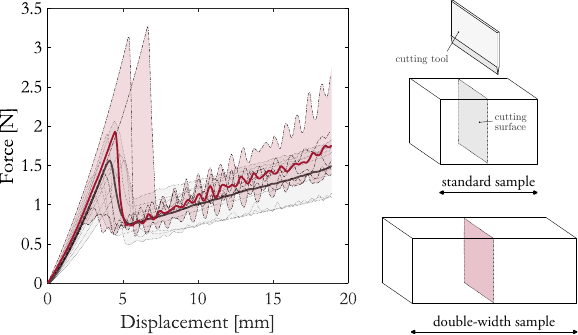}
\caption{\textbf{Additional cutting experiments on gelatin hydrogel samples with double width dimension and same cutting length.} As revealed by the \textit{in silico} framework in the main text, tangential forces at the tool-material interface are primarily governed by adhesive interactions rather than classical Coulomb friction. This finding is supported by experimental results showing that the slope of the cutting force during steady-state cutting remains nearly unchanged when doubling the sample width from \qty{30}{\milli\meter} to \qty{60}{\milli\meter}. The insensitivity of the cutting response to lateral bulk volume suggests that steady-state cutting is predominantly controlled by local interfacial mechanisms rather than global material dimensions.
}
\label{fig:additional_BCs_exp}
\end{figure}

\bibliographystyle{naturemag}%{unsrt}%{ieeetr}
\newpage

\addcontentsline{toc}{section}{References}
%\bibliography{bibliography}

\end{document}